\documentclass[ijoc,nonblindrev]{informs3} % current default for manuscript submission
\usepackage{subfigure}
\usepackage{algorithm}
\usepackage{algorithmic}
\usepackage{cite}
\newtheorem{definition}{Definition}
\newtheorem{lemma}{Lemma}
\newtheorem{theorem}{Theorem}
\usepackage{natbib}
 \bibpunct[, ]{(}{)}{,}{a}{}{,}%
 \def\bibfont{\small}%
 \def\newblock{\ }%
\begin{document}
%%%%%%%%%%%%%%%%

%\setcounter{page}{1}
%\DOI{10.1287/xxxx.0000.0000}%


%\RUNTITLE{}

% \TITLE{Bundling Information Goods of Decreasing Value}
\TITLE{Providing Probabilistic Robustness Guarantee for Crowdsensing}

\ARTICLEAUTHORS{%
\AUTHOR{Yuben Qu$^\dag$, Shaojie~Tang$^\ddag$, Chao~Dong$^\dag$, Peng~Li$^\#$, Song~Guo$^\#$, Chang~Tian$^\dag$}
\AFF{$^\dag$College of Communications Engineering, PLAUST, China}
\AFF{$^\ddag$Naveen Jindal School of Management, University of Texas at Dallas, USA}
\AFF{$^\#$School of Computer Science and Engineering, The University of Aizu, Japan}
} % end of the block

\ABSTRACT{%
Due to its flexible and pervasive sensing ability, crowdsensing has been extensively studied recently in research communities. However, the fundamental issue of how to meet the requirement of sensing robustness in crowdsensing remains largely unsolved. Specifically, from the task owner's perspective, how to minimize the total payment in crowdsensing while guaranteeing the sensing data quality is a critical issue to be resolved. We elegantly model the robustness requirement over sensing data quality as chance constraints, and investigate both hard and soft chance constraints for different crowdsensing applications. For the former, we reformulate the problem through Boole's Inequality, and explore the optimal value gap between the original problem and the reformulated problem. For the latter, we study a serial of a general payment minimization problem, and propose a binary search algorithm that achieves both feasibility and low payment. The performance gap between our solution and the optimal solution is also theoretically analyzed. Extensive simulations validate our theoretical analysis.
% Enter your abstract
}%

%\KEYWORDS{deterministic inventory theory; infinite linear programming duality;
%  existence of optimal policies; semi-Markov decision process; cyclic schedule}


\maketitle
%%%%%%%%%%%%%%%%%%%%%%%%%%%%%%%%%%%%%%%%%%%%%%%%%%%%%%%%%%%%%%%%%%%%%%

% Samples of sectioning (and labeling) in IJOC
% NOTE: (1) \section and \subsection do NOT end with a period
%       (2) \subsubsection and lower need end punctuation
%       (3) capitalization is as shown (title style).
%
%\section{Introduction.}\label{intro} %%1.
%\subsection{Duality and the Classical EOQ Problem.}\label{class-EOQ} %% 1.1.
%\subsection{Outline.}\label{outline1} %% 1.2.
%\subsubsection{Cyclic Schedules for the General Deterministic SMDP.}
%  \label{cyclic-schedules} %% 1.2.1
%\section{Problem Description.}\label{problemdescription} %% 2.

% Text of your paper here
\section{Introduction}
%Crowdsensing is important
In recent years, the provision of real-time environmental information (e.g., traffic condition, noise level and air quality, etc) to citizens is increasingly demanded in daily life. Due to the proliferation of hand-held mobile devices equipped with various sensors, \textit{crowdsensing} \citep{CS} becomes an effective way to collect sensing information with a low deployment cost. In crowdsensing, instead of deploying many fixed sensors in the target area, human crowds with mobile devices (say participants) act as mobile sensors to gather the information of the surrounding area, and utilize the cellular network to upload data for crowdsensing applications. Due to its high efficiency and ubiquity \citep{TTU}, crowdsensing has spurred a wide interest from both academic and industry for designing many interesting crowdsensing applications \citep{MPS}.

%Sensing data quality is crucial for crowdsensing
To fully exploit the benefits of crowdsensing, we need to address several challenges. First, a certain level of sensing robustness should be guaranteed. In practice, it is hardly to accurately estimate how many people will participate in the crowdsensing task, because their decisions are affected by various factors. Since most of involved participants are unprofessional, the sensing data from a single participant would be untrusted or with low quality \citep{OTD,AHA}. Therefore, a minimum number of participants is required to guarantee the sensing robustness, which is imposed by many crowdsensing applications \citep{IMF,TOA}. Moreover, since a crowdsensing task usually involves the collection of sensing data at different times and different locations, e.g., updating the PM2.5 information every hour in the daytime in various areas of a city, the robustness requirement can be both time-specific and location-specific. Then, how to satisfy the time-specific and location-specific requirement is urgent for sensing robustness in crowdsensing.

%The relationship between sensing data quality and minimum participants requirement
Second, since participating in crowdsensing consumes physical resources and requires manual efforts, the participants with rationality will expect financial or social rewards from the \textit{owner} who disseminates the crowdsensing task. Due to limited budget, from the perspective of the owner, it should \textit{minimize the total payment while satisfying the minimum participants requirement as much as possible.} There is a tradeoff between the minimization of total payment and satisfaction of minimum participants requirement. A higher reward (referred as bid in this paper) usually attracts more participants because of a larger accept-to-participate probability from the potential participants \citep{ORB}. This can lead to a high total payment, although the minimum participants requirement can be satisfied easily. On the contrary, if the owner provides a lower bid, people could have less interest in participating resulting in fewer participants, despite a low total payment.

%Challenges
In this paper, we study the problem of minimizing the total payment over all time slots and locations while guaranteeing the sensing robustness in crowdsensing. Different from existing studies, we model the requirement on the sensing robustness as a chance constraint, which guarantees that the probability of achieving the minimum participants requirement at any time slot and any location is not lower than a predefined level. In addition, for many crowdsensing applications, e.g., environment monitoring, imposing such strong requirement of sensing robustness at each time slot is unnecessary. It is sufficient to guarantee a high probability that there are enough participants at any location during the task duration. For example, people sometimes are interested in average air condition during a day. Above models are flexible and well capture the features of sensing robustness requirement in crowdsensing.

%How to solve this problem and our contributions
Bearing in mind different levels of sensing robustness requirement, we propose two types of chance constraints, i.e., hard chance constraint and soft chance constraint. In the hard chance constraint case, we require that the joint probability of achieving the minimum participants requirement at all time slots and locations is no less than $1 - \epsilon$, where $\epsilon > 0$ is close to 0. A typical example is monitoring outages of public works (e.g., broken traffic lights, malfunctioning fire hydrants) by crowdsensing \citep{CS}. In the soft chance constraint case, for any location $l$, the probability of achieving $\alpha_l$ percentage of required participant number during task duration is no less than $\beta$, where $\beta > 0$ is close to 1. It is suitable for the applications with relatively loose robustness requirement, e.g., pollution monitoring in a city \citep{DAC}.

However, solving problems with chance constraints are usually challenging because these constraints are generally difficult to be expressed in a closed form except in very few cases \citep{CAO}. Although some chance-constrained optimization techniques have been applied in existing works \citep[e.g.][]{DRS,RAF,OCR,NCM}, only feasible solutions without performance guarantee are obtained. In this work, to solve the problem with a hard chance constraint, we reformulate it as a solvable convex problem using Boole's Inequality \citep{Bonferroni}, which is an approximation problem of the original one. Their theoretical performance gap is also derived. For the soft chance constraint case, we first transform it into a general payment minimization problem, and analyze the feasibility of its optimal solution to the original problem. Thus motivated, we propose a binary search algorithm to find a feasible solution of the original problem, whose performance gap is theoretically analyzed.

Our main contributions are summarized as follows:
\begin{itemize}
      \item We study the problem of minimizing the total payment with the sensing robustness requirement in crowdsensing, which is yet seldom studied in many crowdsensing applications. Instead of using the expectations in existing works, we are the first to model the payment minimization crowdsensing problem using chance constraints, which better characterize the demand of crowdsensing and is more flexible in realistic. According to the different levels of the sensing robustness requirement, we propose both hard chance constraint and soft chance constraint to model the payment minimization crowdsensing problem.
      \item For the hard chance constraint case, via Boole's Inequality, we reformulate the problem into a solvable convex problem, which is actually an approximation of the original problem. By analyzing the difference in the constraints, we theoretically derive the optimal value gap between the two aforementioned problems, which is largely ignored in previous chance-constrained optimization approaches. Further, by introducing a relaxation problem of the original one, we enhance the above optimal value gap.
      \item For the soft chance constraint case, which cannot be directly reformulated via Boole's Inequality, we study a general payment minimization problem with parameters $\gamma$. By studying the properties of the optimal solutions of this problem, we find that the feasibility of its optimal solution to the original problem strictly increases with the increase of $\gamma$. We then propose a binary search algorithm based on Monte Carlo method, to obtain a feasible solution of the original problem with a high probability, whose performance compared with the optimal solution is also theoretically analyzed. 
      \item We also study a special case of the soft chance constraint in which both the sensing robustness requirement and bidding function are time independent, and derive a closed form of the solution by approximation. The bidding policy from our solution is \textit{state independent}, which is described by a static bidding price (location-specific) over the time. Although the general optimal bidding policy in this kind of problem is \textit{state dependent}, the distinct advantage of our state independent bidding policy is simple to compute, compared with the complex computational optimal policy. The bound of the performance gap between our solution and the optimal solution is also derived.
\end{itemize}

%Paper organization
The rest of this paper is organized as follows. The related works is briefly reviewed in Section 2 and the basic crowdsensing model is introduced in Section 3. We tackle the crowdsensing problem with a hard chance constraint and with multiple soft chance constraints in Section 4 and Section 5, respectively. In Section 6, we conduct extensive simulations to validate our theoretic analysis. Finally, Section 7 concludes the paper. %Introduction
\section{Related Work}
In essence, our aim is to design incentive policies for crowdsensing, to ensure the sensing robustness while minimizing the total payment. There are extensive studies about the incentive design for crowdsensing in existing works.  \citet{EMF} studies the participatory sensing performance of incentive policies with different micro-payment incentive structures. This work mainly focuses on recording the participation likelihood, and pays little attention in incentives. Additionally, there are several empirical experiments \citep{FIA,EMS,ACO} demonstrating the impact of social and financial incentives on the willingness of participants as well as the uploaded sensing data quality. However, these experiment works fail to design a flexible incentive policy to fulfill the diverse sensing robustness requirement.

Additionly, \citet{SYX} propose a reverse auction based dynamic bidding policy to encourage participants to upload the sensing data with the claimed bids. \citet{CTS} proposes incentive policies based on both a user-centric and platform-centric model for crowdsourcing to smartphones, which can adjust the participants' behavior for better crowdsensing performance. \citet{SCI} develops a wireless indoor localization system based on crowdsensing, and guides participants to cover enough locations for the improvement of quality of service. And \citet{OID} studies the problem of minimizing platform's total cost while guaranteeing service quality, and proposes an incentive policy to determine payment allocation and participation level. Moreover, although both \citet{TTU} and \citet{CSP} consider the minimum number of participants in the execution of a crowdsensing task, they simply employ the expectation in the constraint or utilize a deterministic constraint that the requirement should be satisfied at any time. Nonetheless, these studies either have separately addressed the incentive issues and sensing robustness concerns, or have not well characterized the sensing robustness requirement.

In contrast, in this work, we systemically study the incentive policy design with the sensing robustness guarantee in crowdsening, and model the sensing robustness requirement as chance constraints, which are more flexible for designing intelligent incentive policies. Furthermore, we study two kinds of chance constraints based on the different levels of sensing robustness in various crowdsensing applications, i.e., hard chance constraint and soft chance constraint. And in both cases, We theoretically analyze the performance gap between our solution and the optimal solution, while this kind of theoretical results is seldom presented in previous chance-constrained optimization works \citep[e.g.][]{DRS,RAF,OCR,NCM}, which usually obtain feasible solutions only.

%Related work
\section{Model}
\begin{figure}
\FIGURE
{\includegraphics*[scale = 0.43]{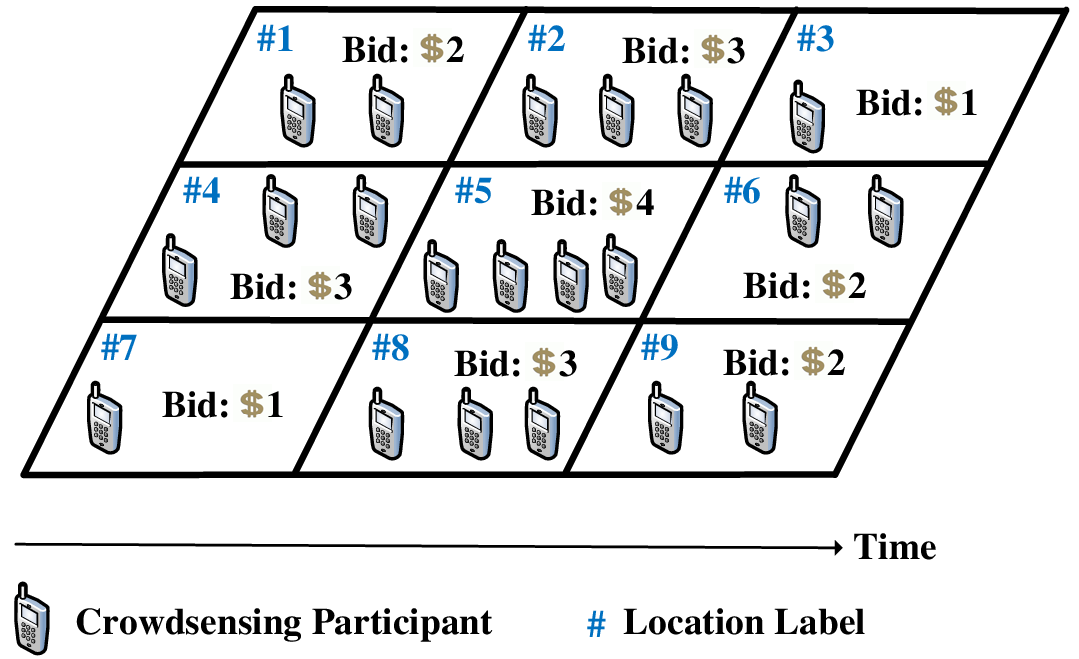}}
{In crowdsensing, the sensing data needs to be collected both at different times and different places.\label{NM}}
{}
\end{figure}

We consider a task owner whose sensing task lasting for a certain amount of time, which is divided into a number of discrete slots, over a target area. Typically, such tasks can be monitoring outages of public works or air pollution for citizens in a city during each day. This type of task usually involves the collection of sensing data at different times and different places, as shown in Fig. \ref{NM}. The sensing robustness requirement of the task may vary in both spatial and temporal dimensions. According to \citep{TTU,CSP}, this requirement can be ensured by involving at least a minimum number of participants in the crowdsensing task.
%\begin{figure}
%\centering
%\includegraphics[scale = 0.43]{NetworkModel}
%\caption{In crowdsensing, the sensing data needs to be collected both at different times and different places.}
%\label{NM}
%\end{figure}
To conduct the task, the owner recruits a number of participants for data collection. In each time slot, the task owner interacts with potential participants as follows:

1) the owner firstly releases the task with its bidding, i.e., the rewards for conducting the task, to all potential participants in the target area;

2) given the bidding, the potential participants decide whether to participate or not, and then send the decision to the owner;

3) upon receiving the response, if the number of potential participants accepting the participation is no less than a predefined threshold (e.g., $r_{min}$), the owner randomly selects $r_{min}$ participants from these potential participants\footnote{How to incorporate the selection of participants with high reputation in our problem is left for future study.}, and pays the corresponding bid price after receiving the sensing data uploaded by these participants; otherwise, the owner won't do anything until next slot, i.e., he will give up the task during the current time slot.

Let $\mathcal{T} = \{1, ..., T\}$ denote the set of time slots. Assume that the target area can be divided into several locations, and the set of locations is denoted by $\mathcal{L} = \{1, ..., L\}$. Let $r_t^l$ and $x_t^l$ be the minimum number of required participants and the actual number of participants chosen by the owner at time $t\in\mathcal{T}$ at location $l\in\mathcal{L}$, respectively. According to the crowdsensing described above, the value of $x_t^l$ can be as follows:
\begin{small}
\begin{align}
&x_t^l = \left\{
     \begin{array}{l l}
           r_t^l & \quad\mbox{if the number of potential participants accepting the} \\
                &  \quad\mbox{bid price at time $t$ at location $l$ is no less than $r_t^l$},\\
           0 & \quad\mbox{otherwise}.\nonumber
     \end{array}\right.
\end{align}
\end{small}

Next, we formally define the strategy profile of the owner and participants respectively as follows:
\begin{definition}(Bidding Vector) The bidding vector of the owner at time $t\in\mathcal{T}$ is defined as $b_t := (b_t^1, ..., b_t^L)$, where $b_t^l$ is the bid for location $l$ and bounded by a maximum value $b_{max}^l$, i.e., $b_t^l\in[0, b_{max}^l]$.
\end{definition}

Note that for different time slots and different locations, the bids may be different. For example, in the traffic volume monitoring application, the amount of the task varies from rush hours to idle hours, as well as from urban areas to suburbs. Given the differentiated amount of the task, a smart owner should also change its bidding policy adaptively. The current bidding vector in Fig. \ref{NM} is (2, 3, 1, 3, 4, 2, 1, 3, 2).

Given a bid, a potential participant randomly decides whether to accept or reject, whose distribution is determined by:
\begin{definition}(Accept-to-Participate Function) The accept-to-participate function of a participant at time $t\in\mathcal{T}$ at location $l\in\mathcal{L}$ is defined by:
\begin{displaymath}\rho_t^l(b): [0, b_{max}^l] \rightarrow [0, 1].\end{displaymath}
\end{definition}

Generally, the greater value of $b$, the larger probability of accepting to participate. However, the increasing trend of the accept-to-participate probability will decrease as the growth of $b$, a.k.a., the marginal effect. In other words, for any $t\in\mathcal{T}, l\in\mathcal{L}$, $\rho_t^l(b)$ is strictly increasing and concave in $b\in[0, b_{max}^l]$ \citep{ORB}.

To facilitate the mathematical formulation of our problem, we define the bidding function $b_t^l(\rho)$ as the inverse function of $\rho_t^l(b)$ as follows.
\begin{definition}(Bidding Function) The bidding function of the owner at time $t\in\mathcal{T}$ at location $l\in\mathcal{L}$ is defined by:
\begin{displaymath}b_t^l(\rho): [0, 1] \rightarrow [0, b_{max}^l].\end{displaymath}
\end{definition}By definition, $b_t^l(\rho)$ is the required bid at time $t$ at location $l$ to ensure an accept-to-participate probability of $\rho$. Since $\rho_t^l(b)$ is strictly increasing and concave in $b\in[0, b_{max}^l]$, $b_t^l(\rho)$ is strictly increasing and convex in $\rho\in[0, 1]$.

Let $A_t^l$ be the event of $x_t^l = r_t^l$ ($\forall t\in\mathcal{T}$, $l\in\mathcal{L}$). If we define $\rho_t^l$ as the success probability that $A_t^l$ happens, i.e., $\rho_t^l = \mbox{Pr}\{x_t^l = r_t^l\}$, then the expected payment at time $t$ at location $l$ will be
\begin{align}
\label{eq1}
\mathbb{E}[x_t^l b_t^l] = \rho_t^l r_t^l\times b_t^l(\rho_t^l),
\end{align}where $\rho_t^l r_t^l$ is the expected number of participants chosen by the owner in the participation at time $t$ at location $l$, and $b_t^l(\rho_t^l)$ is the bid for participants at time $t$ at location $l$ ensuring an accept-to-participate probability $\rho_t^l$ among the potential participants. Therefore, the total expected payment of the owner over all time slots and locations is
\begin{align}
\label{eq2}
\mathbb{E} \left[\sum_{t=1}^T \sum_{l=1}^L x_t^l b_t^l\right] = \sum_{t=1}^T \sum_{l=1}^L \mathbb{E}[x_t^l b_t^l] = \sum_{t=1}^T \sum_{l=1}^L \rho_t^l r_t^l b_t^l(\rho_t^l).
\end{align}

We study this problem from the owner's perspective with the objective of minimizing the total payment over all time slots and locations, while satisfying the sensing robustness requirement as much as possible. Depending on the satisfactory level of the sensing robustness requirement in different crowdsensing applications, we model two types of chance constraints in the optimization problem. i.e., hard chance constraint and soft chance constraint. In the first type of constraint, the joint probability that the sensing robustness requirement at all time slots and locations can be satisfied is no less than $1 - \epsilon$, while in the second one, for any location $l\in\mathcal{L}$, the probability that at least $\alpha_l$ percent of the requirement can be satisfied during the task duration is no less than $\beta$. We deal with the crowdsensing problems with the two types of chance constraint in Section 4 and Section 5, respectively.

%Model
\section{Crowdsensing with Hard Chance Constraint}
In this section, we study the crowdsensing problem with a hard chance constraint on the sensing robustness requirement. Due to the hardness using the chance constraint, this problem will be reformulated using Boole's Inequality, in order to obtain a feasible approximation solution of the original problem. We then derive the performance gap between the optimal solution and the approximation one.
\subsection{Problem Statement}
In essence, the owner's objective is to minimize the total payment, while maintaining the probability of satisfying the sensing robustness requirement at \textit{any time slot} and \textit{any location} no less than a predefined value $1 - \epsilon$, where $\epsilon$ is a small value close to 0, e.g., 0.02.

Based on Eq. (\ref{eq2}) and the definition of $\rho_t^l$, we mathematically formulate the crowdsensing problem with a hard chance constraint in the following:\\
\textbf{PA1: Crowdsensing with a Hard Chance Constraint\footnote{Note that optimizing $p_t^l$ is equivalent to optimizing $b_t^l$, since there is a one-to-one mapping relationship between them in the bidding function.}}
\begin{align}
&\mbox{min} \quad\quad\quad\quad \sum_{t=1}^T \sum_{l=1}^L \rho_t^l r_t^l b_t^l(\rho_t^l) \nonumber\\
\label{eq3}
&\mbox{subject to} \quad \prod_{t=1}^T \prod_{l=1}^L \rho_t^l \geq 1 - \epsilon, \\
&\quad\quad\quad\quad\quad\quad\rho_t^l\in[0, 1], \forall t\in\mathcal{T}, l\in\mathcal{L}. \nonumber
\end{align}

The above problem is hard to solve, due to the combinatorial difficulty in the multiplication form in constraint (\ref{eq3}), where the variable $\rho_t^l$ is continuous. In the following, we reformulate the problem into a solvable convex problem, by using Boole's Inequality. Through this reformulation, the new constraint will share a similar additive structure with the objective function, which enables the problem to be efficiently solved.

\subsection{Reformulation via Boole's Inequality}
Firstly, based on the definition of $\rho_t^l$, we have $\mbox{Pr}\{A_t^l\} = \mbox{Pr}\{x_t^l = r_t^l\} = \rho_t^l$. Then, constraint (\ref{eq3}) is equivalent to
\begin{align}
\label{eq4}
\mbox{Pr}\left\{\bigwedge_{t=1}^T \bigwedge_{l=1}^L A_t^l\right\} \geq 1 - \epsilon.
\end{align}
Also, we define the indicator function $I_t^l(x_t^l)$ as follows
\begin{align}
\label{eq5}
&I_t^l(x_t^l) := \left\{
     \begin{array}{l l}
           0 & \quad\mbox{if $x_t^l = r_t^l$},\\
           1 & \quad\mbox{otherwise}.
     \end{array}\right.
\end{align}
Eq. (\ref{eq5}) means that $I_t^l(x_t^l)$ equals to 1, if $x_t^l$ does not meet the sensing robustness requirement at time $t$ at location $l$. Accordingly, $I_t^l(x_t^l)$ represents the unsatisfiability at time $t$ at location $l$. Via Boole's Inequality \citep{Bonferroni}, we can reformulate PA1 as follows: \\
\textbf{PA2: Approximation to PA1}
\begin{align}
&\mbox{min} \quad\quad\quad\quad \sum_{t=1}^T \sum_{l=1}^L \rho_t^l r_t^l b_t^l(\rho_t^l) \nonumber\\
\label{eq6}
&\mbox{subject to} \quad \mathbb{E}\left\{\sum_{t=1}^T \sum_{l=1}^L I_t^l(x_t^l) \right\} \leq \epsilon.
\end{align}
The following lemma ensures that any feasible solution of PA2 is feasible for PA1.
\begin{lemma}\label{lemma1}Any feasible solution of PA2 is also a feasible solution of PA1.\end{lemma}
\proof{Proof.} Since the objectives of PA1 and PA2 are identical, we only need to prove that (\ref{eq6}) is a sufficient condition for (\ref{eq4}). In fact, according to the definition of $I_t^l(x_t^l)$, the probability that $A_t^l$ does not happen equals to the expectation of $I_t^l(x_t^l)$, i.e., $1 - \mbox{Pr}\{A_t^l\} = \mathbb{E}\{I_t^l(x_t^l) \}.$
Using this equation and Boole's Inequality, i.e., $\mbox{Pr}\{A_1\bigvee \ldots \bigvee A_N\} \leq \sum_{i=1}^N \mbox{Pr}\{A_i\}$, the LHS of (\ref{eq4}) can be lower bounded by
\begin{align}
\label{eq8}
&\mbox{Pr}\left\{\bigwedge_{t=1}^T\bigwedge_{l=1}^L A_t^l\right\} \geq 1 - \mathbb{E}\left\{\sum_{t=1}^T\sum_{l=1}^L I_t^l(x_t^l) \right\}.
\end{align}Therefore, if (\ref{eq6}) holds, (\ref{eq4}) also holds based on the fact of (\ref{eq8}). The lemma is thus proved.\Halmos

\subsection{Performance Gap Analysis}
PA2 can be seen as a conservative approximation to PA1, which substitutes the LHS of (\ref{eq4}) by the RHS of (\ref{eq8}) in (\ref{eq4}). Note that Lemma \ref{lemma1} only gives the feasibility guarantee. However, the approximation gap between the optimal values of PA2 and PA1 is still unclear. In the following, in Section 4.3.1, we first analyze the performance gap between the approximation solution and the optimal solution by analyzing the conservatism in the constraints, and provide an initial bound on the performance gap. And we then introduce a relaxed problem of PA1 to enhance the bound on the performance gap in Section 4.3.2.
\subsubsection{First Bound on the Performance Gap:}
We begin with the analysis of the difference in the conservatism in the approximation of PA2 to PA1 in the following lemma.
\begin{lemma}\label{lemma2}The difference between the LHS and RHS of the inequality in (\ref{eq8}) is no more than $\frac{TL(TL-1)}{2} \epsilon^2$.\end{lemma}

\proof{Proof.}Let $\bar{A}_t^l$ be the complementary event of $A_t^l$. Then the difference between the LHS and RHS of the inequality in (\ref{eq8}), i.e., $\bigtriangleup$, can be rewritten as
\begin{align}
&\bigtriangleup = \mbox{Pr}\left\{\bigwedge_{t=1}^T\bigwedge_{l=1}^L A_t^l\right\} - \left[1 - \mathbb{E}\left\{\sum_{t=1}^T\sum_{l=1}^L I_t^l(x_t^l)\right\} \right] \nonumber\\
\label{eq9}
&\quad= \left[1 - \mbox{Pr}\left\{\bigvee_{t=1}^L \bigvee_{l=1}^L \bar{A}_t^l\right\} \right] - \left[1 - \sum_{t=1}^T\sum_{l=1}^L \mbox{Pr}\{\bar{A}_t^l\} \right] \\
\label{eq10}
&\quad= \sum_{t=1}^T\sum_{l=1}^L \mbox{Pr}\{\bar{A}_t^l\} - \mbox{Pr}\left\{\bigvee_{t=1}^L \bigvee_{l=1}^L \bar{A}_t^l\right\}
\end{align}

From (\ref{eq4}), we know that
$\mbox{Pr}\left\{\bigvee_{t=1}^T\bigvee_{l=1}^L \bar{A}_t^l \right\} \leq \epsilon$, and thus $\mbox{Pr} \{\bar{A}_t^l \} \leq \epsilon$ holds for any $t = 1, ..., T$ and $l = 1, ..., L$. According to Bonferroni Inequalities \citep{Bonferroni}, we have
$0 \leq \bigtriangleup \leq \sum_{t < t'}\sum_{l < l'} \mbox{Pr}\left\{\bar{A}_t^l \bigwedge \bar{A}_{t'}^{l'}\right\}.$
In our problem, since the satisfaction of the sensing robustness requirement is relatively independent at any time slot and location, $\{\bar{A}_t^l\}_{t=1, ..., T, l=1, ..., L}$ is probabilistically independent with each other, then
\begin{align}
\sum_{t < t'}\sum_{l < l'} \mbox{Pr}\left\{\bar{A}_t^l \bigwedge \bar{A}_{t'}^{l'}\right\} \leq \sum_{t < t'}\sum_{l < l'} \mbox{Pr}\left\{\bar{A}_t^l\right\} \mbox{Pr}\{\bar{A}_{t'}^{l'}\} \leq \sum_{t < t'}\sum_{l < l'} \epsilon\cdot\epsilon = \frac{TL(TL-1)}{2} \epsilon^2. \nonumber
\end{align}The lemma thus holds.\Halmos

Next, we will analyze the performance gap between the optimal solution of PA2 and that of PA1 based on the above conservatism result. To simplify the notations, we let \\$F(\rho) := \sum_{t=1}^T \sum_{l=1}^L \rho_t^l r_t^l b_t^l(\rho_t^l)$, $G(\rho) := 1 - \prod_{t=1}^T \prod_{l=1}^L \rho_t^l$, and $H(\rho) := \sum_{t=1}^T \sum_{l=1}^L (1 - \rho_t^l)$, where $\rho := \{\rho_t^l\}_{t\in\mathcal{T}, l\in\mathcal{L}}$, $\rho_t^l\in[0, 1], \forall t\in\mathcal{T}, l\in\mathcal{L}$. Thus, PA1 and PA2 can be equivalently represented as $\mbox{min}_{_{G(\rho) \leq \epsilon}} F(\rho)$, and $\mbox{min}_{_{H(\rho) \leq \epsilon}} F(\rho)$, respectively. To obtain the performance gap, we first have the following lemma.
\begin{lemma}\label{lemma3}Let $\tilde{\lambda}$ and $\tilde{\rho}$ be the optimal solutions of
\begin{align}
\label{eq13}
\max_{\lambda\geq 0} \min_{\rho} \left\{F(\rho) + \lambda[H(\rho) - \epsilon] \right\}.
\end{align}
Thus, $\tilde{\rho}$ is an optimal solution of PA2.
\end{lemma}

\proof{Proof.}We prove this lemma by showing that PA2 is a convex problem with strong duality. For the convexity, the inequality function $H(\rho)$ is obviously convex, according to the linear combination of $\rho_t^l$ in $H(\rho)$. As to the objective function, we only need to prove the convexity of $\rho_t^lb_t^l(\rho_t^l)$. Since both $b_t^l(\rho_t^l)$ and $\rho_t^l$ are strictly increasing and convex in $\rho_t^l\in[0, 1]$, their product function $\rho_t^lb_t^l(\rho_t^l)$ is convex \citep{Convex}.

Moreover, $H(\rho)$ represents the expected total infeasibility levels over all time slots and locations under a bidding sequence $\{b_t^l(\rho_t^l)\}_{t\in\mathcal{T}, l\in\mathcal{L}}$. Thus, if we release large enough bids at each time slot and location without considering the payment, the number of potential participants that accept to participate will be always enough. In this case, $H(\rho)$ equals to 0. This means that there exists a variable $\rho$ (corresponding to the bids $\{b_t^l\}_{t\in\mathcal{T}, l\in\mathcal{L}}$) that satisfies $H(\rho) = 0 < \epsilon$. In other words, the Slater's condition holds for PA2. Combining with the convex property, strong duality holds for PA2. Therefore, $\tilde{\rho}$ is a primal optimal solution of PA2 \citep{Convex}. The lemma thus holds.\Halmos

Based on Lemma \ref{lemma2} and Lemma \ref{lemma3}, we can obtain the initial bound about the optimal value gap between PA1 and PA2 in the following theorem.

\begin{theorem}\label{theorem1}Let $\rho^\ast$ be an optimal solution of PA1. And let $\tilde{\lambda}$ and $\tilde{\rho}$ be the optimal solutions of (\ref{eq13}), then the gap between the optimal value of PA1 and that of PA2 can be bounded by
\begin{equation}
\label{eq17}
0 \leq F(\tilde{\rho}) - F(\rho^\ast) \leq \tilde{\lambda}\frac{TL(TL - 1)}{2} \epsilon^2.
\end{equation}
\end{theorem}

\proof{Proof.}
We derive the bound in two steps. According to Lemma \ref{lemma3}, $\tilde{\rho}$ is an optimal solution of PA2. We begin with the definitions of $\tilde{\lambda}$ and $\tilde{\rho}$, which are the solutions of (\ref{eq13}). We then have
\begin{align}
&F(\tilde{\rho}) = \max_{\lambda\geq 0} \min_{\rho} \left\{F(\rho) + \lambda [H(\rho) - \epsilon]\right\} \nonumber \\
&\quad\quad= \min_{\rho} \left\{F(\rho) + \tilde{\lambda} [H(\rho) - \epsilon]\right\} \nonumber \\
&\quad\quad\leq F(\rho^\ast) + \tilde{\lambda} [H(\rho^\ast) - \epsilon] \nonumber \\
&\quad\quad= F(\rho^\ast) + \tilde{\lambda} [G(\rho^\ast) - \epsilon + H(\rho^\ast) - G(\rho^\ast)] \nonumber \\
&\quad\quad= F(\rho^\ast) + \tilde{\lambda} (G(\rho^\ast) - \epsilon) + \tilde{\lambda} [H(\rho^\ast) - G(\rho^\ast)] \nonumber \\
&\quad\quad\leq F(\rho^\ast) + \tilde{\lambda} [H(\rho^\ast) - G(\rho^\ast)] \nonumber \\
\label{eq18}
&\quad\quad\leq F(\rho^\ast) + \tilde{\lambda} \frac{TL(TL - 1)}{2} \epsilon^2.
\end{align}The second equality follows the definition of $\tilde{\lambda}$ and the strong duality property. The first inequality is due to the definition of $\tilde{\rho}$. The second inequality is because of $\tilde{\lambda} \geq 0$ and $G(\rho^\ast) \leq \epsilon$. And the last inequality follows Lemma \ref{lemma2}. Further, since the objective is a minimization and Lemma \ref{lemma1} holds, we have
\begin{align}
\label{eq19}
F(\rho^\ast) \leq F(\tilde{\rho}).
\end{align}Combining (\ref{eq18}) and (\ref{eq19}), the theorem thus holds.\Halmos

\subsubsection{Second Bound by Introducing a Relaxation:}
Theorem \ref{theorem1} shows that the performance gap between the optimal solution of PA1 and that of PA2 is actually of the order of $O(T^2L^2\epsilon^2)$. In the following, we construct a relaxed problem of PA1 to provide a possible tighter bound. Recall that in Lemma \ref{lemma2}, we have obtained $\mbox{Pr} \{\bar{A}_t^l \} \leq \epsilon$ for $\forall t\in\mathcal{T}, l\in\mathcal{L}$, which obviously implies $\sum_{t=1}^T \sum_{l=1}^L \mbox{Pr} \{\bar{A}_t^l \} \leq TL\epsilon$. Therefore, substituting $\epsilon$ by $TL\epsilon$ in (\ref{eq6}), we actually get a \textit{relaxed} problem of PA1 in the following:\\
\textbf{PA3: Relaxation to PA1}\\
\begin{align}
&\mbox{min} \quad\quad\quad\quad \sum_{t=1}^T \sum_{l=1}^L \rho_t^l r_t^l b_t^l(\rho_t^l) \nonumber\\
\label{eq15}
&\mbox{subject to} \quad \mathbb{E}\left\{\sum_{t=1}^T \sum_{l=1}^L I_t^l(x_t^l) \right\} \leq TL\epsilon.
\end{align}Similar to PA1 and PA2, PA3 can be simplified as $\mbox{min}_{_{H(\rho) \leq TL\epsilon}} F(\rho).$

Since PA3 is a relaxed problem of PA1, for any feasible solution $\rho$ of PA1, $\rho$ is also a feasible solution of PA3. Furthermore, utilizing PA3, the performance gap between the optimal solution of PA1 and that of PA2 can also be bounded by the following theorem.

\begin{theorem}\label{lemma4}Let $\rho^\ast$ be an optimal solution of PA1. And let $\tilde{\lambda}$ and $\tilde{\rho}$ be the optimal solutions of (\ref{eq13}). Thus, the gap between the optimal value of PA1 and that of PA2 can be also bounded by
\begin{equation}
\label{eq17}
0 \leq F(\tilde{\rho}) - F(\rho^\ast) \leq \tilde{\lambda} (TL - 1) \epsilon.
\end{equation}
\end{theorem}

\proof{Proof.}Due to Lemma \ref{lemma3}, $\tilde{\lambda}$ is obviously an optimal solution of PA2. Recall that PA3 is a relaxation to PA1, and the objective is identical as $\min_{\rho} F(\rho)$. Let $\hat{\rho}$ be an optimal solution of PA3. Combining the result of (\ref{eq19}), we have
\begin{align}
\label{eq14}
F(\hat{\rho}) \leq F(\rho^\ast) \leq F(\tilde{\rho}),
\end{align}which implies
\begin{align}
\label{eq21}
0 \leq F(\tilde{\rho}) - F(\rho^\ast) \leq F(\tilde{\rho}) - F(\hat{\rho}).
\end{align}

Note that the constraint in PA3 is equivalent to $\frac{1}{TL}H(\rho) \leq \epsilon.$ Similar to the proof in (\ref{eq18}), we have
\begin{align}
&F(\tilde{\rho}) \leq F(\hat{\rho}) + \tilde{\lambda} [H(\hat{\rho}) - \frac{1}{TL}H(\hat{\rho})] \nonumber \\
&\quad\quad= F(\hat{\rho}) + \tilde{\lambda} \frac{TL - 1}{TL} H(\hat{\rho}) \nonumber \\
\label{eq20}
&\quad\quad\leq F(\hat{\rho}) + \tilde{\lambda} (TL - 1) \epsilon,
\end{align}where the last inequality is because of $H(\hat{\rho}) \leq TL\epsilon$.

Due to (\ref{eq21}) and (\ref{eq20}), we have
%\begin{align}
%\label{eq22}
$0 \leq F(\tilde{\rho}) - F(\rho^\ast) \leq \tilde{\lambda} (TL - 1) \epsilon.$
%\end{align}
The proof is thus concluded.\Halmos

Combining the results of Theorem \ref{theorem1} and Theorem \ref{lemma4}, we formally record the bound about the performance gap between the optimal solution of PA1 and that of PA2 below.
\begin{theorem}\label{theorem2}
Let $\rho^\ast$ be an optimal solution of PA1. And let $\tilde{\lambda}$ and $\tilde{\rho}$ be the optimal solutions of (\ref{eq13}). Then, the gap between the optimal value of PA1 and that of PA2 can be bounded by
$0 \leq F(\tilde{\rho}) - F(\rho^\ast) \leq \tilde{\lambda} \min\left\{(TL - 1) \epsilon, \frac{TL(TL - 1)}{2} \epsilon^2\right\}.$
\end{theorem}

In Theorem \ref{theorem2}, if $TL > \frac{2}{\epsilon}$, $0 \leq F(\tilde{\rho}) - F(\rho^\ast) \leq \tilde{\lambda} (TL - 1)\epsilon$; otherwise, $0 \leq F(\tilde{\rho}) - F(\rho^\ast) \leq \tilde{\lambda} \frac{TL(TL - 1)}{2} \epsilon^2$. Theorem \ref{theorem2} shows the bound about the performance gap between the optimal solution of the reformulated problem (PA2) and that of the original problem (PA1). Since both $F(\rho)$ and $H(\rho)$ are differentiable, we can easily obtain the value of $\tilde{\lambda}$ by the KKT optimality condition \citep{Convex}, i.e., $\nabla F(\tilde{\rho}) + \tilde{\lambda} \nabla H(\tilde{\rho}) = 0$. If $H(\tilde{\rho}) < 0$, we then have $\tilde{\lambda} = 0$; otherwise, we can bound $\tilde{\lambda}$ by $\tilde{\lambda} \leq \frac{\|\nabla F(\tilde{\rho})\|}{\|\nabla H(\tilde{\rho})\|}.$

%Crowdsensing with joint chance constraint
\section{Crowdsensing with Soft Chance Constraint}
Different from Section 4, we model the constraint of the sensing robustness requirement as multiple \textit{soft chance constraints} in this section. Specifically, the soft chance constraints ensure that for any location $l\in\mathcal{L}$, the robustness requirement is satisfied at least $\alpha_l$ ($\alpha_l\in(0, 1]$) percentage over $T$ time slots with a probability of $\beta$ or more, where $\beta$ is close to 1, e.g., 98\%. Compared to the hard chance constraint (\ref{eq3}) in PA1 in Section 4, this constraint can not only be more suitable for some kind of tasks with a slightly loose requirement, but also better characterize the location-specific requirement in the task execution.
\subsection{Problem Statement}
To formulate the crowdsensing problem with soft chance constraints, we first define an indicator function $\bar{I}_t^l(x_t^l)$, which is the opposite of $I_t^l(x_t^l)$ in Eq. (\ref{eq5}), i.e.,
\begin{align}
\label{eq24}
&\bar{I}_t^l(x_t^l) := \left\{
     \begin{array}{l l}
           1 & \quad\mbox{if $x_t^l = r_t^l$},\\
           0 & \quad\mbox{otherwise}.
     \end{array}\right.
\end{align}Accordingly, we know that $\mathbb{E}\{\bar{I}_t^l\} = \rho_t^l$. In light of Eq. (\ref{eq24}), the average times that the sensing robustness requirement is satisfied at location $l\in\mathcal{L}$ over task duration $T$ is $\sum_{t=1}^T \bar{I}_t^l(x_t^l)/T$.

The crowdsensing problem with multiple soft chance constraints can be mathematically formulated as follows:\\
\textbf{PB1: Crowdsensing with Multiple Soft Chance Constraints}
\begin{align}
&\mbox{min} \quad\quad\quad\quad \sum_{t=1}^T \sum_{l=1}^L \rho_t^l r_t^l b_t^l(\rho_t^l) \nonumber\\
\label{eq25}
&\mbox{subject to} \quad \mbox{Pr}\left\{\frac{\sum_{t=1}^T \bar{I}_t^l(x_t^l)}{T} \geq \alpha_l\right\} \geq \beta, \forall l\in\mathcal{L},\\
&\quad\quad\quad\quad\quad\quad\rho_t^l\in[0, 1], \forall t\in\mathcal{T}, l\in\mathcal{L}. \nonumber
\end{align}Similar to PA1 in Section 4, PB1 is also untractable due to the chance constraint in (\ref{eq25}). Unfortunately, we cannot directly apply a similar approximation approach in Section 4 to PB1, because of the more challenging form of (\ref{eq25}).

To obtain a feasible solution of PB1, we study a general payment minimization problem related to PB1, by transforming (\ref{eq25}) into a constraint \textit{in expectation} for the total times of success requirement satisfaction for any location as follows:\\
\textbf{PB2: A General Payment Minimization Problem with $\{\gamma_l\}_{l\in\mathcal{L}}$}
\begin{align}
&\mbox{min} \quad\quad\quad\quad \sum_{t=1}^T \sum_{l=1}^L \rho_t^l r_t^l b_t^l(\rho_t^l) \nonumber\\
\label{eq26}
&\mbox{subject to} \quad \sum_{t=1}^T \rho_t^l \geq \gamma_l, \forall l\in\mathcal{L}, \\
&\quad\quad\quad\quad\quad\quad\rho_t^l\in[0, 1], \forall t\in\mathcal{T}, l\in\mathcal{L}, \nonumber
\end{align}which utilizes the fact
\begin{displaymath}\mathbb{E} \left\{\sum_{t=1}^T  \bar{I}_t^l(x_t^l)\right\} = \sum_{t=1}^T  \mathbb{E} \left\{\bar{I}_t^l(x_t^l)\right\} = \sum_{t=1}^T \rho_t^l, \forall l\in\mathcal{L}.\end{displaymath}

In PB1, (\ref{eq25}) guarantees that the probability on the success percentage for each location $l\in\mathcal{L}$ is lower bounded by $\beta$, i.e., \begin{displaymath}\mbox{Pr}\left\{\frac{\sum_{t=1}^T \bar{I}_t^l(x_t^l)}{T} \geq \alpha_l\right\} \geq \beta,\end{displaymath} which can imply
\begin{align}
\label{eq27}
&\mathbb{E}\left\{\frac{\sum_{t=1}^T \bar{I}_t^l(x_t^l)}{T}\right\} \geq \alpha_l \beta \Longrightarrow \sum_{t=1}^T \rho_t^l \geq T\alpha_l\beta.
\end{align}As a result, if we set $\gamma_l = T\alpha_l\beta$ in PB2, we can obtain a relaxed problem of PB1 as follows: \\
\textbf{PB3: Relaxation to PB1}
\begin{align}
&\mbox{min} \quad\quad\quad\quad \sum_{t=1}^T \sum_{l=1}^L \rho_t^l r_t^l b_t^l(\rho_t^l) \nonumber\\
\label{eq28}
&\mbox{subject to} \quad \sum_{t=1}^T \rho_t^l \geq T\alpha_l\beta, \forall l\in\mathcal{L}, \\
&\quad\quad\quad\quad\quad\quad\rho_t^l\in[0, 1], \forall t\in\mathcal{T}, l\in\mathcal{L}. \nonumber
\end{align}Although PB3 can be solved by classical convex approaches, its feasible solution may not be feasible for PB1, since PB3 is only a relaxed version of PB1. Fortunately, in this work, we find that PB1 and PB2 have a very appealing relationship between their solutions.

We begin with some important notations. Let $h^l(\rho^l)$ be the LHS of (\ref{eq25}) in PB1 for each $l\in\mathcal{L}$, i.e.,
\begin{align}
\label{eq29}
h^l(\rho^l) = \mbox{Pr}\left\{\frac{\sum_{t=1}^T \bar{I}_t^l(x_t^l)}{T} \geq \alpha_l\right\},
\end{align}where $\rho^l := \{\rho_t^l\}_{t\in\mathcal{T}}$. And denote the optimal solution of PB2 corresponding to $\gamma$ be $\rho^\ast(\gamma) = \{{\rho^\ast}^l(\gamma_l)\}_{l\in\mathcal{L}}$, where $\gamma := \{\gamma_l\}_{l\in\mathcal{L}}$. Next, we study how the value of $h^l({\rho^\ast}^l(\gamma_l))$ will change with the variation of $\gamma_l$. Our findings will reveal that, if we increase the value of $\gamma_l$, $h^l({\rho^\ast}^l(\gamma_l))$ also increases. In other words, $h^l({\rho^\ast}^l(\gamma_l))$ has a positive correlation with the value of $\gamma_l$ in fact. This motivates us to design an algorithm that can find a value of $\tilde{\gamma}$, so as to obtain a solution $\rho^\ast(\tilde{\gamma})$ feasible for PB1 and with low payment.

In the rest of this section, we will analyze the properties of optimal solutions of PB2 in Section 5.2, and theoretically prove the above observations in the end of this subsection. Based on the results of Section 5.2, we then propose a binary search algorithm based on Monte Carlo method that finds a solution feasible for PB1 with a high probability in Section 5.3. Using a very similar analysis method as in Section 4, we also theoretically derive the performance gap between our solution and the optimal solution. Last, in Section 5.4, we study a special case of PB1 by assuming that both the requirement $r_t^l$ and bidding function $b_t^l(\cdot)$ are time independent, and derive a closed form solution by approximation.
\subsection{Properties of Optimal Solutions of PB2}
In this part, we explore several interesting and useful properties of the optimal solutions of PB2. We begin with some necessary notations. For simplicity, we first let $f_{tl}(\cdot)$ be the inside of the summation in the objective function in PB2, i.e., $f_{tl}(\rho_t^l) = \rho_t^l r_t^l b_t^l(\rho_t^l), \forall t\in\mathcal{T}, l\in\mathcal{L}.$ Let $\rho^l := \{\rho_t^l\}_{t = 1, ..., T}$ and $F^l(\rho^l) := \sum_{t=1}^T f_t^l(\rho_t^l).$ Thus, the objective function can also be represented as $F(\rho) = \sum_{l=1}^L F^l(\rho^l)$, where $\rho = \{\rho^l\}_{l = 1, ..., L}$. Recall that in Lemma \ref{lemma3}, we have verified that $f_{tl}(\rho_t^l)$ is strictly increasing and convex in $\rho_t^l\in[0, 1]$, for any $t\in\mathcal{T}, l\in\mathcal{L}$, which implies
\begin{align}
\label{eq31}
f'_{tl}(\rho_t^l) > 0, f''_{tl}(\rho_t^l) \geq 0, \forall \rho_t^l\in[0, 1].
\end{align}

Furthermore, since $b_t^l(\cdot)$ is strictly increasing and convex, and $f''_{tl}(\rho_t^l) = r_t^l[2{b_t^l}'(\rho_t^l) + \rho_t^l {b_t^l}''(\rho_t^l)]$, we have
\begin{align}
\label{eq32}
f''_{tl}(\rho_t^l) > 0, \forall \rho_t^l\in[0, 1],
\end{align}based on ${b_t^l}'(\rho_t^l) > 0$ and ${b_t^l}''(\rho_t^l) \geq 0$ for $\forall \rho_t^l\in[0, 1]$.

Since the objective function has a separable structure in location $l\in\mathcal{L}$, and the constraints for each location $l$ are actually independent with each other, solving PB2 is equivalent to solving $L$ independent subproblems as follows:\\
\textbf{PB2.$l$: The $l$th Subproblem of PB2}
\begin{align}
&\mbox{min} \quad\quad\quad\quad \sum_{t=1}^T f_{tl}(\rho_t^l) \nonumber\\
\label{eq30}
&\mbox{subject to} \quad \sum_{t=1}^T \rho_t^l \geq \gamma_l, \rho_t^l\in[0, 1], \forall t\in\mathcal{T}.
%&\quad\quad\quad\quad\quad\quad\rho_t^l\in[0, 1], \forall t\in\mathcal{T}. \nonumber
\end{align}Suppose that ${\rho^\ast}^l = \{{\rho_t^\ast}^l\}_{t = 1, ..., T}$ is an optimal solution of PB2.$l$ ($l = 1, ..., L$), then $\rho^\ast = \{{\rho^\ast}^l\}_{l = 1, ..., L}$ will be an optimal solution of PB2; and vice versa. Therefore, in the following, we focus on the $l$th subproblem only, whose properties in the optimality also hold for any other subproblems.

Since $\rho_t^l\in[0, 1]$ for $\forall t\in\mathcal{T}, l\in\mathcal{L}$, if $\gamma_l > T$, PB2.$l$ will be unsolvable. If $\gamma_l = T$, the corresponding optimal solution will be meaningless in practical, i.e., ${\rho^\ast_t}^l = 1$, $t = 1, ..., T$, which corresponds to the policy that always releases the largest bid, although it is surely feasible for PB2. Also, if $\gamma_l \leq 0$, ${\rho^\ast_t}^l = 0$, $t = 1, ..., T$, which is obviously infeasible for PB2. As a result, the rest of this section is based on the assumption of $0 < \gamma_l < T$ for $l = 1, ..., L$ unless specified.

For the first derivative of any optimal solution ${\rho^\ast}^l = \{{\rho_t^\ast}^l\}_{t = 1, ..., T}$ for PB2.$l$, we have the following lemma.
\begin{lemma}\label{lemma5}Let ${\rho^\ast}^l := \{{\rho_t^\ast}^l\}_{t = 1, ..., T}$ be an optimal solution of PB2.$l$, and $f'_{tl}({\rho_t^\ast}^l)$ be the first derivative of function $f_{tl}(\cdot)$ at ${\rho_t^\ast}^l$, $t = 1, ..., T$. Then, we can conclude that
\begin{displaymath}
f'_{1l}({\rho_1^\ast}^l) = f'_{2l}({\rho_2^\ast}^l) = \cdot\cdot\cdot = f'_{Tl}(\rho_T^{\ast l}).
\end{displaymath}
\end{lemma}

\proof{Proof.}
The proof is provided in Appendix A.\Halmos

Lemma \ref{lemma5} reveals a very interesting property in the property of optimal solutions of PB2.$l$, i.e., the first derivatives of each $f_{tl}(\cdot)$ ($t = 1, ..., T$) at any optimal solution are equal. This nice property can be used to show how the value of the optimal solution will change with the parameter $\gamma_l$, as illustrated in the following lemma.

\begin{lemma}\label{lemma6}Let ${\rho^\ast}^l(\gamma_l) := \{{\rho_t^\ast}^l(\gamma_l)\}_{t = 1, ..., T}$ and ${\rho^\ast}^l(\hat{\gamma_l}) := \{{\rho_t^\ast}^l(\hat{\gamma_l})\}_{t = 1, ..., T}$ be the optimal solutions corresponding to $\gamma_l$ and $\hat{\gamma_l}$ in (\ref{eq30}) for PB2.$l$, respectively. If $\hat{\gamma_l} > \gamma_l$, then for $\forall t = 1, ..., T$, we have ${\rho_t^\ast}^l(\hat{\gamma_l}) > {\rho_t^\ast}^l(\gamma_l).$
\end{lemma}

\proof{Proof.}
The proof is provided in Appendix B.\Halmos

Lemma \ref{lemma6} reveals another important property of the optimal solution of PB2.$l$: if we increase the value of $\gamma_l$, the value of the corresponding optimal solution (${\rho^\ast}^l(\gamma_l) := \{{\rho_t^\ast}^l(\gamma_l)\}_{t = 1, ..., T}$) also strictly increases, i.e., ${\rho_t^\ast}^l(\gamma_l)$ is strictly increasing in $\gamma_l\in(0, T)$ for any $t = 1, ..., T$. Based on this property, we are able to demonstrate how $h^l({\rho^\ast}^l(\gamma_l))$, which is the LHS of (\ref{eq25}), will change with the increase of $\gamma_l$ in the following theorem.

\begin{theorem}\label{theorem3}Let ${\rho^\ast}^l(\gamma_l) := \{{\rho_t^\ast}^l(\gamma_l)\}_{t = 1, ..., T}$ be an optimal solution of PB2.$l$ with $\gamma_l$. Then, $h^l({\rho^\ast}^l(\gamma_l))$ as defined in (\ref{eq29}) is strictly increasing with the increase of $\gamma_l$.
\end{theorem}

\proof{Proof.}
The proof can be found in Appendix C.\Halmos

According to Theorem \ref{theorem3}, if we increase the value of $\gamma_l$, the probability of $h^l({\rho^\ast}^l(\gamma_l)) \geq \beta$ is also increasing, since $h^l({\rho^\ast}^l(\gamma_l))$ increases. This means that the probability that $\rho^{\ast l}(\gamma_l) := \{\rho^{\ast l}_t(\gamma_l)\}_{t = 1, ..., T}$ satisfies the constraint (\ref{eq30}) in PB2.$l$ will be higher with the increase of $\gamma_l$. Thanks to this observation, we are able to find the smallest $\gamma_l$ that returns a feasible solution of PB1, which is introduced in detail in the following subsection.

\subsection{A Binary Search Algorithm for PB1}
In this subsection, we propose a binary search algorithm for PB1 based on Monte Carlo method, to obtain a feasible bidding policy for the crowdsensing problem with soft chance constraints. The \textit{key insight} of our algorithm is as follows. When we increase the value of $\gamma_l$ from 0 for each subproblem of PB2, the feasibility of its optimal solution to PB1 improves. Since the objective of PB2 is identical to that of PB1, this optimal solution can be a good approximation solution of PB1 if some feasibility condition holds. Note that in our algorithm, we utilize binary search to find an appropriate $\overline{\gamma_l}$ between the initial interval $[0, T]$. And the feasibility condition is checked by Monte Carlo method. In the following, we introduce the proposed algorithm in detail.

\begin{algorithm}
\caption{Binary Search Algorithm for Crowdsensing with Soft Chance Constraints}
\label{alg1}
\begin{algorithmic}
\STATE{\textbf{Input:} Error tolerance $\overline{\sigma} > \underline{\sigma} > 0$}
\STATE{\textbf{Output:} Bidding policy $b^{\overline{\gamma}}$ for PB1}
\STATE{\textbf{Initialization:} For $l = 1, ..., L$, set $\underline{\gamma_l} = 0$ and $\overline{\gamma_l} = T$.}
\STATE{1: \textbf{for} $l = 1, ..., L$}
\STATE{2: \quad\textbf{while} $q^l(\overline{\gamma_l}) > \overline{\sigma} \parallel q^l(\overline{\gamma_l}) < \underline{\sigma}$ \textbf{do}}
\STATE{3: \quad\quad\quad Obtain the optimal solution of PB2.$l$ with $\gamma_l = \frac{\overline{\gamma_l} + \underline{\gamma_l}}{2}$, i.e., $\rho^{\ast l} = \{\rho^{\ast l}_t\}_{t = 1, ..., T}$}
\STATE{4: \quad\quad\quad Compute $h^l(\rho^{\ast l})$ by Monte Carlo method}
\STATE{5: \quad\quad\quad\textbf{if} $q^l(\frac{\overline{\gamma_l} + \underline{\gamma_l}}{2}) < 0$ \textbf{then}}
\STATE{6: \quad\quad\quad\quad\quad $\underline{\gamma_l} \leftarrow \frac{\overline{\gamma_l} + \underline{\gamma_l}}{2}$}
\STATE{7: \quad\quad\quad\textbf{else}}
\STATE{8: \quad\quad\quad\quad\quad $\overline{\gamma_l} \leftarrow \frac{\overline{\gamma_l} + \underline{\gamma_l}}{2}$}
\STATE{9: \quad\quad\quad\textbf{end if}}
\STATE{10: \quad\textbf{end while}}
\STATE{11: \quad Record the optimal solution of PB2.$l$ with $\overline{\gamma_l}$, e.g., $\rho^{\ast l}(\overline{\gamma_l}) = \{\rho^{\ast l}_t(\overline{\gamma_l})\}_{t = 1, ..., T}$, and \\ \qquad\quad then calculate the corresponding bidding prices, i.e., \begin{displaymath}b^{\ast l}(\overline{\gamma_l}) = \{b_t^l(\rho^{\ast l}_t(\overline{\gamma_l}))\}_{t = 1, ..., T}\end{displaymath}}
\STATE{12: \textbf{end for}}
\STATE{13: \textbf{return} $b^{\overline{\gamma}} = \{b^{\ast l}(\overline{\gamma_l})\}_{l = 1, ..., L}$}
\end{algorithmic}
\end{algorithm}

For convenience, we first define $g^l(\rho^l) := h^l(\rho^l) - \beta, l = 1, ..., L,$ where $h^l(\rho^l)$ is defined in Eq. (\ref{eq29}). By definition, a sequence $\rho := \{\rho^l\}_{l = 1, ..., L}$ is feasible for PB1 if $g^l(\rho^l)$ is not negative, for any $l = 1, ..., L$. Let $\rho^{\ast l}(\gamma_l)$ be the optimal solution of PB2.$l$ with $\gamma_l$. For simplicity, we denote $g^l(\rho^{\ast l}(\gamma_l))$ by $q^l(\gamma_l)$ in the algorithm, for $l = 1, ..., L$.

As shown in Algorithm 1, we first initialize the search interval $[\underline{\gamma_l}, \overline{\gamma_l}]$ as [0, T], for each $l = 1, ..., L$. Then, from Line 1, we independently search the appropriate $\overline{\gamma_l}$ for PB2.$l$ from $l = 1$ to $L$. Line 2 provides the terminal condition for each $l$, i.e., whenever $\underline{\sigma} \leq q^l(\overline{\gamma_l}) \leq \overline{\sigma}$ holds, it records the optimal solution corresponding to $\overline{\sigma}$, and then calculates the corresponding bidding prices based on the bidding function, as illustrated in Line 11. Note that the terminal condition means that, we have found an appropriate value $\overline{\gamma_l}$, whose corresponding optimal solution of PB2.$l$ ($\rho^{\ast l}(\overline{\gamma_l})$) is feasible for the $l$th subproblem of PB1 with a high probability, and the expected payment is not large. The former property is ensured by imposing a lower bound on $q^l(\overline{\gamma_l})$, i.e., $q^l(\overline{\gamma_l}) \geq \underline{\sigma}$, which is used to reduce the inevitable error introduced by Monte Carlo method. The latter property is guaranteed by upper bounding $q^l(\overline{\gamma_l})$ to a value slightly larger than $\underline{\sigma}$, i.e., $q^l(\overline{\gamma_l}) \leq \overline{\sigma}$.

Lines 3 - 9 show the main loop of Algorithm 1. In Line 3, the optimal solution of PB2.$l$ with $\gamma_l$, which is the middle value of the current interval, is obtained, i.e., $\rho^{\ast l} = \{\rho^{\ast l}_t\}_{t = 1, ..., T}$. Then, in Line 4, the value of $h^l(\rho^{\ast l})$ is computed by Monte Carlo method. Specifically, consider an experiment of $T$ independent events, whose success probabilities are $\{\rho^{\ast l}_t\}_{t = 1, ..., T}$, respectively. Repeat the experiment $N$ times, and record the number of success events in the $n$th experiment as $k_n$. Then, $h^l(\rho^{\ast l})$ can be approximated as $\frac{\sum_{n=1}^N \mathbf{1}(k_n \geq k_l)}{N}$, where $\mathbf{1}(\cdot)$ is the indicator function. In Line 5, we check whether $q^l(\frac{\overline{\gamma_l} + \underline{\gamma_l}}{2})$ is negative or not. If yes, we update the lower value of the search interval as the middle value in Line 6, i.e., $\underline{\gamma_l} = \frac{\overline{\gamma_l} + \underline{\gamma_l}}{2}$. Otherwise, the upper value of the interval is updated in Line 8, i.e., $\overline{\gamma_l} = \frac{\overline{\gamma_l} + \underline{\gamma_l}}{2}$. When the search for all $L$ subproblems has been completed, we return the bidding policy that contains the bidding prices in all subproblems as the output.

Last, we analyze the performance of the bidding policy computed by Algorithm 1, and provide an upper bound on the gap between the total payment by the proposed bidding policy and the optimal payment. Similar as in the hard chance constraint case, our \textit{key idea} is to introduce a conservative approximation problem of PB1 and then derive the optimal value gap between this approximation problem and PB3 (a relaxed problem of PB1), which can bound the performance gap between the optimal solution and the solution computed by Algorithm 1. We have the following theorem:

\begin{theorem}\label{theorem6}
Let $F(\rho^{\ast}(\overline{\gamma}))$ and $F^\ast$ denote the payment with $\rho^{\ast}(\overline{\gamma})$ obtained by Algorithm 1 and the optimal payment of PB1, respectively. Then, the difference between $F(\rho^{\ast}(\overline{\gamma}))$ and $F^\ast$ can be bounded by
\begin{align}
\label{eq64}
0 \leq F(\rho^{\ast}(\overline{\gamma})) - F^\ast \leq \sum_{l=1}^L \lambda^\ast_l(T-T\alpha_l\beta-1+\beta), \nonumber
\end{align}where $\lambda^\ast_l$ is the optimal solution of the $l$th dual problem
\begin{displaymath}
\max_{\lambda_l \geq 0} \min_{0 \leq \rho_t^l \leq 1} \left\{\sum_{t=1}^T \rho_t^l r^l b^l(\rho_t^l) + \lambda_l(T - 1 + \beta - \sum_{t=1}^T \rho_t^l)\right\}
\end{displaymath}for $l = 1, ..., L$.
\end{theorem}

\proof{Proof.}The proof mainly contains two steps. First, by setting the required percentage $\alpha_l$ as 1 in PB1 and using a similar approximation of PA2 to PA1 in Section 4, we obtain a very conservative approximation problem of PB1 as follows:\\
\textbf{PB4: Approximation to PB1}
\begin{align}
&\mbox{min} \quad\quad\quad\quad \sum_{t=1}^T \sum_{l=1}^L \rho_t^l r_t^l b_t^l(\rho_t^l) \nonumber\\
\label{eq62}
&\mbox{subject to} \quad \sum_{t=1}^T \rho_t^l \geq T - 1 + \beta, \forall l\in\mathcal{L}, \\
&\quad\quad\quad\quad\quad\quad\rho_t^l\in[0, 1], \forall t\in\mathcal{T}, l\in\mathcal{L}. \nonumber
\end{align}Let $F^\ast_a$ and $F^\ast_r$ are the optimal values of PB4 and PB3, respectively. Notice that both PB3 and PB4 can be solved by independently solving $L$ subproblems (e.g., PB3.$l$ and PB4.$l$, $l = 1, ..., L$), where both PB3.$l$ and PB4.$l$ have a similar expression with PB2.$l$. And the difference between the constraint in PB3.$l$ and that in PB4.$l$ is no more than $T - T\alpha_l\beta - 1 + \beta$. Then, applying the same process in Theorem \ref{theorem1}, the optimal value gap between PB3.$l$ and PB4.$l$ can be bounded by $\lambda^\ast_l(T - T\alpha_l\beta - 1 + \beta)$, where $\lambda^\ast_l$ is the optimal solution of the following dual problem
\begin{displaymath}
\max_{\lambda_l \geq 0} \min_{0 \leq \rho_t^l \leq 1} \left\{\sum_{t=1}^T \rho_t^l r^l b^l(\rho_t^l) + \lambda_l(T - 1 + \beta - \sum_{t=1}^T \rho_t^l)\right\}.
\end{displaymath}Summing the gaps between each subproblems, we easily have that the optimal value gap between PB3 and PB4 is upper bounded by $F^\ast_a - F^\ast_r \leq \sum_{l=1}^L \lambda^\ast_l(T-T\alpha_l\beta-1+\beta).$

Second, let $F(\rho^{\ast}(\overline{\gamma}))$ and $F^\ast$ denote the payment with $\rho^{\ast}(\overline{\gamma})$ (obtained by Algorithm 1) and the optimal payment of PB1, respectively. According to the relationship among the three problems (i.e., PB1, PB3 and PB4), we then have
\begin{displaymath}
F^\ast_r \leq F^\ast \leq F(\rho^{\ast}(\overline{\gamma})) \leq F^\ast_a,
\end{displaymath}which implies $0 \leq F(\rho^{\ast}(\overline{\gamma})) - F^\ast \leq F^\ast_a - F^\ast_r.$ Based on this inequality, the proof is thus concluded.\Halmos

\subsection{A Special Case of PB1}
In this part, we consider a special case of PB1, in which both the sensing robustness requirement $r_t^l$ and the  bidding function $b_t^l(\cdot)$ for any fixed location $l\in\mathcal{L}$ are time independent, i.e., for each location $l\in\mathcal{L}$, $r_t^l = r^l, \forall t\in\mathcal{T}$ and $b_t^l(\cdot) = b^l(\cdot)$. In other words, for any location $l\in\mathcal{L}$, a same bidding price $b$ at different times can generate a exactly same accept-to-anticipate probability among the potential participants. Generally, the optimal bidding policy of this problem is \textit{state dependent}, whose computational complexity will be huge when the task duration $T$ is very large. In this case, however, we are able to obtain a bidding policy that is \textit{state independent} and simple to compute, based on the time independent property of the requirement and bidding function. Specifically, a state independent solution $\rho = \{\rho_t^l\}_{t\in\mathcal{T}, l\in\mathcal{L}}$ for PB1 will have the following property:
\begin{align}
\label{eq43}
\rho_t^l = \rho^l, \forall t\in\mathcal{T},
\end{align}for each $l\in\mathcal{L}$, which can make constraint (\ref{eq25}) tight in PB1. That is,
\begin{align}
\label{eq44}
\mbox{Pr} \left\{\frac{K_{T+1}^l}{T} \geq \alpha_l\right\} \geq \beta, \forall l\in\mathcal{L},
\end{align}where $K_{T+1}^l$ is the sum of success times in satisfying the sensing data quality requirement during the whole duration $T$, should hold.

Since the bidding policy corresponding to $\rho$ is static (time independent), the random variable $K_{T+1}^l$ is binomial distributed, i.e., $K_{T+1}^l \sim \mbox{Bin}(T, \tilde{\rho}^l), l\in\mathcal{L}$, where $\tilde{\rho}^l$ is the trial probability that the sensing data quality requirement is successfully satisfied. Let $\Omega_{B(T, \tilde{\rho}^l})$ represent the CDF of the binomial distribution. In the following, we will derive the value of $\tilde{\rho}^l$ that satisfies (\ref{eq43}) by approximation, for each $l\in\mathcal{L}$. According to (\ref{eq43}), we have
\begin{align}
\label{eq44}
\bar{\Omega}_{B(T, \tilde{\rho}^l)}(T\alpha_l) = \beta, \forall l\in\mathcal{L},
\end{align}where $\bar{\Omega}_{B(T, \tilde{\rho}^l)}(T\alpha_l) = 1 - \Omega_{B(T, \tilde{\rho}^l)}(T\alpha_l)$. To obtain a closed form of $\tilde{\rho}^l$, we use the Normal distribution to approximate $B(T, \tilde{\rho}^l)$ here, since the task duration $T$ can be very large in the soft chance constraint case. Thus, we have
\begin{displaymath}
\bar{\Omega}_N\left(\frac{T\tilde{\rho}^l - T\alpha_l}{\sqrt{T\tilde{\rho}^l(1 - \tilde{\rho}^l)}}\right) = \beta, \forall l\in\mathcal{L},
\end{displaymath}where $\Omega_N(\cdot)$ is the CDF of the Standard Normal distribution. By some operations, we can approximate $\tilde{\rho}^l$ as
\begin{displaymath}
\tilde{\rho}^l \approx \alpha_l + x_{\beta}\sqrt{\frac{\alpha_l}{T}}, \forall l\in\mathcal{L},
\end{displaymath}where $x_{\beta} = \Omega_N^{-1}(\beta)$.

With this approximation solution, the expected percentage in the success satisfaction of the sensing requirement will be $\alpha_l + x_{\beta}\sqrt{\frac{\alpha_l}{T}}$, which is strictly larger than $\alpha_l$, for any $l\in\mathcal{L}$. Then, we can construct a solution $\rho = \{\rho^l\}_{l=1, ..., L} = \{\rho_t^l\}_{t=1, ..., T, l=1, ..., L}$ as follows:
\begin{align}
\label{eq49}
\rho_t^l = \alpha_l + x_{\beta}\sqrt{\frac{\alpha_l}{T}}, \forall t=1, ..., T,
\end{align}for each $l = 1, ..., L$.

By construction, it is not difficult to verify that $\rho$ is feasible for PB1. Next, we analyze the gap between the optimal payment of PB1 and the total payment with $\rho$. We first introduce a \textit{deterministic} counterpart of PB2 as follows: \\
\textbf{PB5: Deterministic Counterpart of PB2}
\begin{align}
\label{eq45}
&\mbox{min} \quad\quad\quad\quad T \sum_{l=1}^L \hat{\rho}^l r^l b_l(\hat{\rho}^l) \\
\label{eq46}
&\mbox{subject to} \quad T\hat{\rho}^l \geq \gamma_l, \forall l\in\mathcal{L}, \\
&\quad\quad\quad\quad\quad\quad\hat{\rho}^l\in[0, 1], \forall l\in\mathcal{L}. \nonumber
\end{align}For convenience, we let $\hat{F}(\hat{\rho})$ denote the objective function of PB5, where $\hat{\rho} := \{\hat{\rho}^l\}_{l\in\mathcal{L}}$. In the relationship between the optimal solutions of PB2 and PB5, we have the following lemma.

\begin{lemma}\label{lemma7}
Let $\hat{\rho}^{\ast} = \{\hat{\rho}^l_{\ast}\}_{l\in\mathcal{L}}$ be an optimal solution of PB5. Then, $\rho^{\ast} = \{{\rho_t^l}^{\ast}\}_{t\in\mathcal{T}, l\in\mathcal{L}}$ will be an optimal solution of PB2, where ${\rho_t^l}^{\ast} = \hat{\rho}^l_\ast, \forall t\in\mathcal{T}$, for each $l\in\mathcal{L}$.
\end{lemma}

\proof{Proof.} The proof is provided in Appendix D.\Halmos

Setting $\gamma_l = T\alpha_l + x_{\beta}\sqrt{T\alpha_l}$, the solution $\hat{\rho} = \{\rho^l\}_{l=1, ..., L}$ will be an optimal solution of PB5, where $\rho^l = \alpha_l + x_{\beta}\sqrt{\frac{\alpha_l}{T}}, l = 1, ..., L$. According to Lemma \ref{lemma7}, $\rho$ as defined in Eq. (\ref{eq49}) is an optimal solution of PB2. Since PB3 is a relaxation problem of PB1, we can now employ the similar method to analyze the optimal value gap between PB3 and PB5 with $\gamma_l = T\alpha_l + x_{\beta}\sqrt{T\alpha_l}$, which is a upper bound for the gap between the optimal payment of PB1 and the total payment with the solution $\rho$.

\begin{theorem}\label{theorem5}
Define $\rho$ as in Eq. (\ref{eq49}). Let $F^\ast$ be the optimal payment of PB1. Then, the gap between $F^\ast$ and the total payment with $\rho$ is bounded by
\begin{displaymath}
0 \leq F(\rho) - F^\ast \leq \sum_{l=1}^L \hat{\lambda}_l(\gamma_l - T\alpha_l\beta),
\end{displaymath}where $\hat{\lambda}_l$ is the optimal solution of the $l$th dual problem
\begin{displaymath}
\max_{\lambda_l \geq 0} \min_{0 \leq \rho_t^l \leq 1} \left\{\sum_{t=1}^T \rho_t^l r^l b^l(\rho_t^l) + \lambda_l(\gamma_l - \sum_{t=1}^T \rho_t^l)\right\}
\end{displaymath}for $l = 1, ..., L$.
\end{theorem}

\proof{Proof.} (\textit{Sketch}) The gap between the $F^\ast$ and the total payment with $\rho$ is no more than the optimal value gap between PB3 and PB5 with $\gamma_l = T\alpha_l + x_{\beta}\sqrt{T\alpha_l}$ ($l = 1, ..., L$). Notice that both PB3 and PB5 can be solved by solving $L$ subproblems independently. Applying the same process in Theorem \ref{theorem1}, we can obtain the optimal value gap between each subproblem. Summing the gaps during each subproblem, we then have the final gap.\Halmos

%Crowdsensing with probabilitic cosntraint
\section{Performance Evaluation}
In this section, we perform simulations in a crowdsensing scenario with Matlab to validate our theoretic analysis. Specifically, we compare our bidding policy with the three following policies: 1) the bidding policy derived from the optimal solution of PA3 for the hard chance constraint case and PB3 for the soft chance constraint case, respectively, which can be seen as an lower bound for the optimal payment of our problem in both cases (called as ``\textit{Lower Bound}'' in the following); 2) an uniform policy that chooses the bidding price from the bidding set with probability $1 - \epsilon$ in the hard case and with probability $\beta$ in the soft case (labeled as ``\textit{Uniform Policy}'' in the figures); 3) a random bidding policy that randomly selects the bidding price from the bidding set during each time slot for each location (labeled as ``\textit{Random Policy}''). For the performance metrics, we mainly use the time-average difference of total payment between a policy (Our Policy/Uniform Policy/Random Policy) and the policy corresponding to Lower Bound, and the probability that the sensing robustness requirement is satisfied in the simulations.\footnote{Note that this metric is not employed in the soft chance case, since this probability has been ensured by $\overline{\sigma}$ and $\underline{\sigma}$ in Algorithm 1.}

\subsection{Setup}
We consider a common crowdsensing scenario that needs sensing a target area over one week with 10 hours in each day. For instance, an environment NGO intends to collect the air quality information for citizens in a city, i.e., updating the indexes including PM2.5 every hour during daytime. Defining one hour as a time slot, the number of time slots is $T = 70$. And the target area can be divided into $L = 6$ different locations that coverage six critical sensing regions. The number of required participants at time $t$ at location $l$, i.e., $r_t^l$, is randomly chosen from the interval $[1, l^2]$. The bidding function at time $t$ at location $l$ is set as $b_t^l(x) = lx^3$, which is identical for any time $t$ at a fixed location $l$. As to the parameter setting, we vary the parameter $\epsilon$ from 0 to 0.08 with a step size 0.02 in the hard chance constraint case, while varying $\beta$ from 0.91 to 0.99 with a step size 0.02 in the soft chance constraint case. Moreover, in the latter case, for the required success percentage at location $l$ (i.e., $\alpha_l$), we consider two settings, i.e., $\alpha_l$ is randomly chosen from $[0.9, 1]$ or from $[0.75, 1]$. In the simulation setting of Algorithm 1, we let $\overline{\sigma}$ and $\underline{\sigma}$ be 0.02 and 0.01, respectively. In the Monte Carlo simulations, we set the number of repeated simulations in an experiment as 500.

\subsection{Results}

\subsubsection{Hard Chance Constraint Case:}
\begin{figure}
\FIGURE
{\includegraphics*[scale = 0.3]{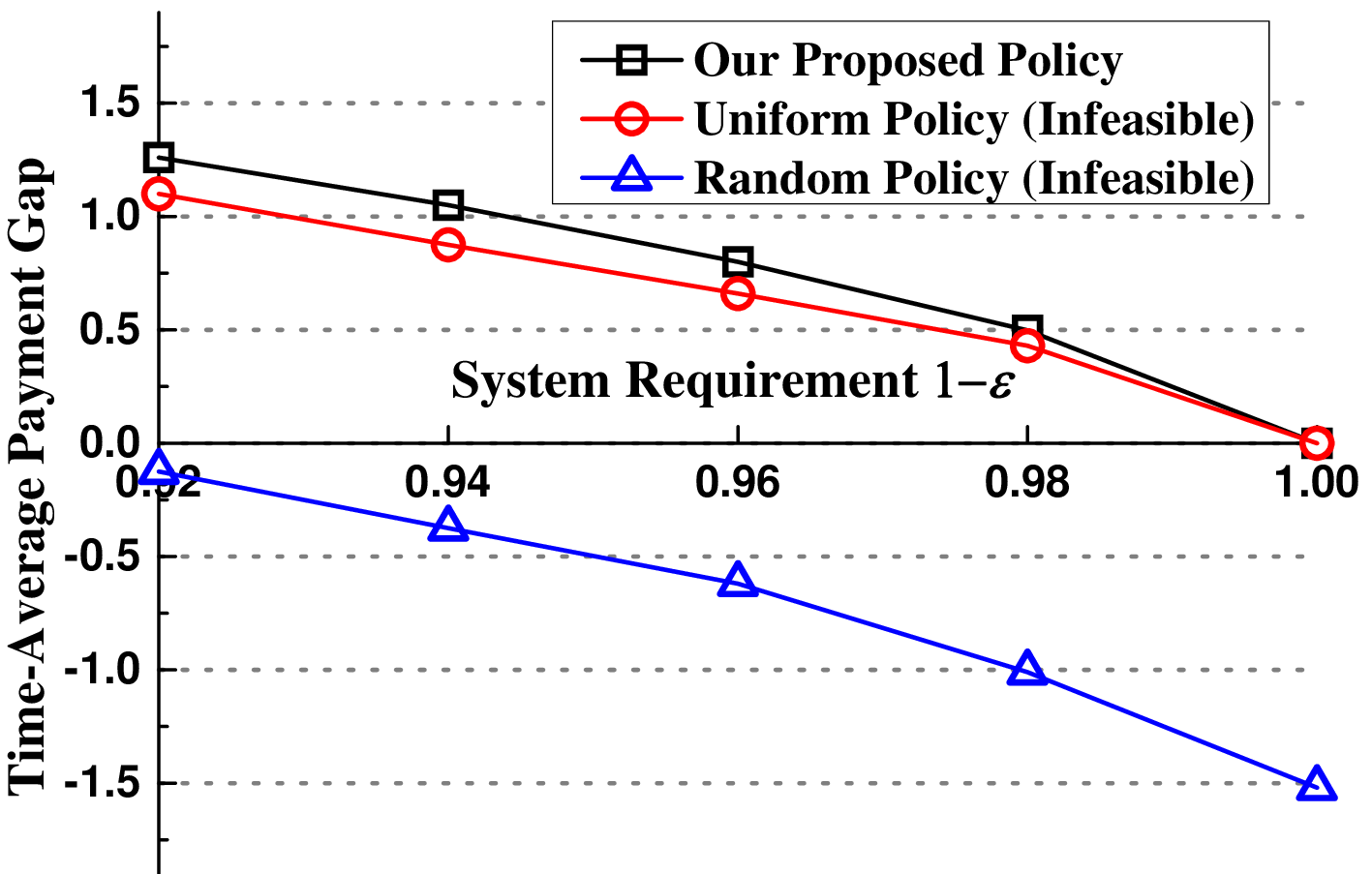}}
{Time-Average Payment Gap vs. System Requirement.\label{HCC_Payment}}
{}
\end{figure}

We first show the simulation results for the hard chance constraint case. Fig. \ref{HCC_Payment} illustrates the time-average gap of total payment between each bidding policy and the bidding policy corresponding to Lower Bound. From the figure, the time-average gap of total payment of both our proposed policy and uniform policy decreases with the increase of the system requirement $1-\epsilon$. This is because when $1-\epsilon$ increases, corresponding to the case with more stringent system requirement on the sensing robustness, both policies as well as the ``Lower Bound'' policy will provides large bids to satisfy the sensing robustness requirement. In contrary, there is no awareness of the sensing robustness requirement in the random policy, which results in a lower payment than the ``Lower Bound'' policy. And we find that the time-average gap of total payment between our proposed policy and  the ``Lower Bound'' policy is not large, especially under stringent sensing robustness requirement.

%\begin{figure}[!htb]
%\centering
%\includegraphics[scale = 0.22]{HCC_Payment_Gap}
%\caption{Time-Average Gap vs. System Requirement.}
%\label{HCC_Payment}
%\end{figure}

Table 1 shows the results of the joint probability that the sensing robustness requirement at any time and location is satisfied under the four bidding policies, compared with the system requirement. From the table, we can clearly obtain that our policy can always meet the system requirement, while all the other policies are \textit{infeasible}. Furthermore, to reduce the total payment, the joint probability under our policy is just slightly larger than the system requirement.
%\begin{table}[!hbtp]
%\centering
%\scriptsize
%\label{TAB1}
%\caption{Success Prob. vs. System Requirement}
%\begin{tabular}{|c|c|c|c|c|}%Ë«À¸ÓÃ
%\begin{tabular}{|l|p{3.5in}|}
%  \hline
%  $1-\epsilon$ & Our Policy & Lower Bound & Uniform Policy & Random Policy \\
%  \hline
%  0.92 & 0.9229 & 0.6855 & 0.0204 & 1.03E-14 \\
%  \hline
%  0.94 & 0.9416 & 0.7524 & 0.0557 & 5.62E-15 \\
%  \hline
%  0.96 & 0.9607 & 0.8304 & 0.1488 & 5.46E-15 \\
%  \hline
%  0.98 & 0.9802 & 0.9234 & 0.3895 & 1.06E-15 \\
%  \hline
%  1 & 1 & 1 & 1 & 6.24E-15 \\
%  \hline
%\end{tabular}
%\end{table}

\begin{table}
\TABLE
{Success Prob. vs. System Requirement (Hard Chance Constraint).\label{TAB1}}
{\begin{tabular}{|c|c|c|c|c|}%Ë«À¸ÓÃ
  \hline
  $1-\epsilon$ & Our Policy & Lower Bound & Uniform Policy & Random Policy \\
  \hline
  0.92 & 0.9229 & 0.6855 & 0.0204 & 1.03E-14 \\
  \hline
  0.94 & 0.9416 & 0.7524 & 0.0557 & 5.62E-15 \\
  \hline
  0.96 & 0.9607 & 0.8304 & 0.1488 & 5.46E-15 \\
  \hline
  0.98 & 0.9802 & 0.9234 & 0.3895 & 1.06E-15 \\
  \hline
  1 & 1 & 1 & 1 & 6.24E-15 \\
  \hline
\end{tabular}}
{}
\end{table}
\subsubsection{Soft Chance Constraint Case:}
\begin{figure}
\FIGURE
{\includegraphics*[scale = 0.3]{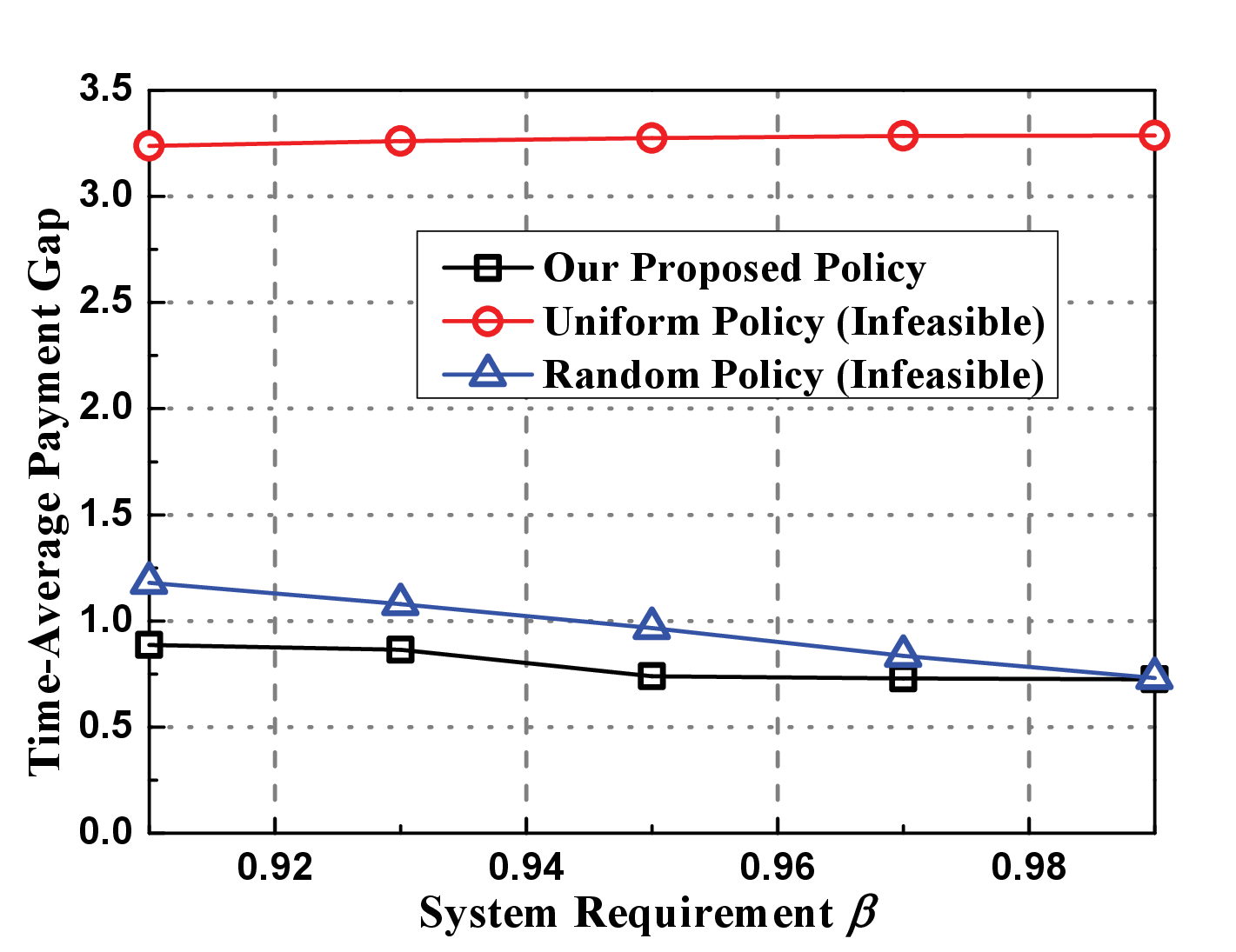}}
{Time-Average Payment Gap vs. System Requirement under Setting I.\label{SCC_Payment_1}}
{}
\end{figure}

\begin{figure}
\FIGURE
{\includegraphics*[scale = 0.3]{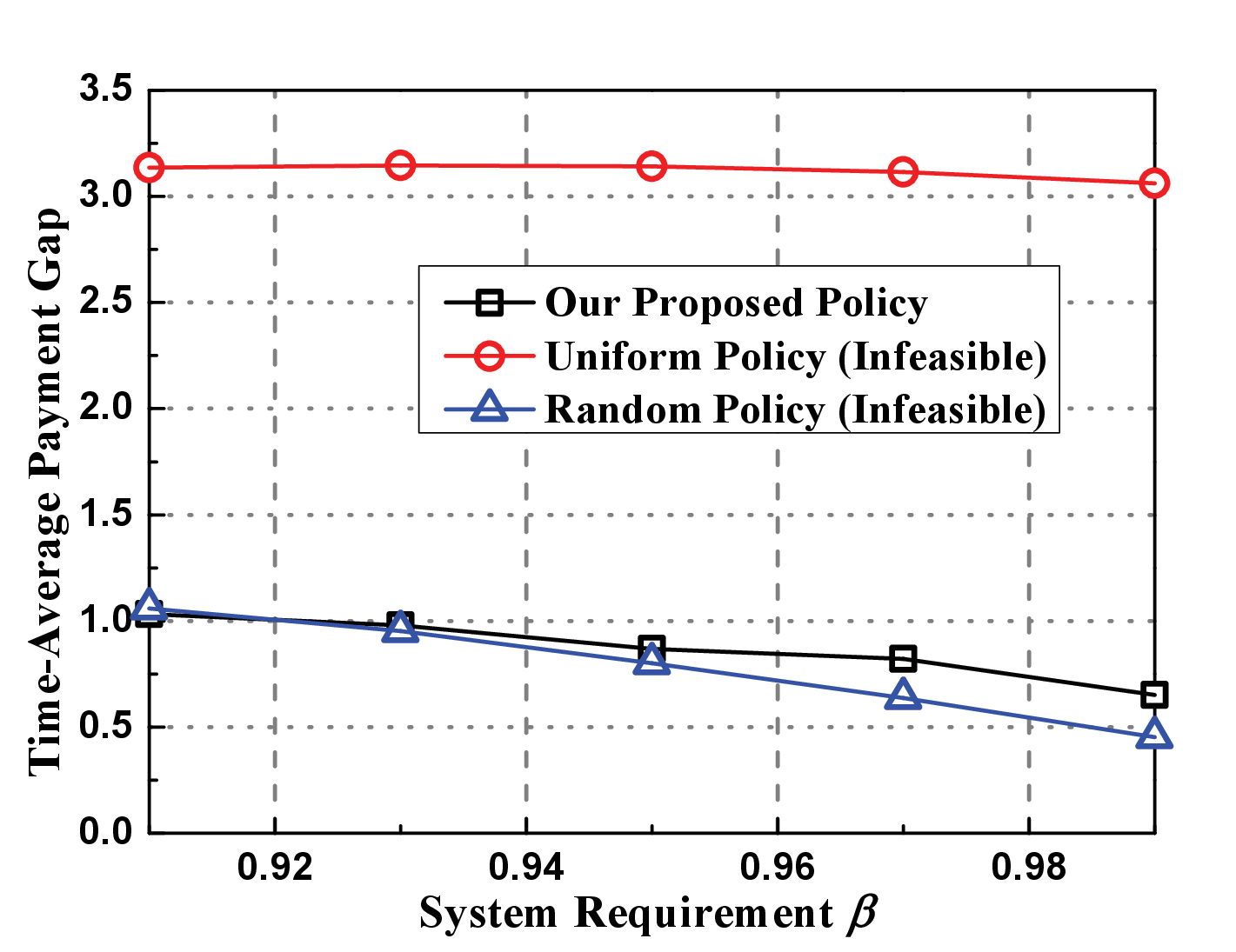}}
{Time-Average Payment Gap vs. System Requirement under Setting II.\label{SCC_Payment_2}}
{}
\end{figure}

Next, in Figs. \ref{SCC_Payment_1} and \ref{SCC_Payment_2}, we study the time-average gap of total payment of our bidding policy computed by Algorithm 1, compared to the uniform policy and random policy. First, according to Fig. \ref{SCC_Payment_1}, our policy achieves the lowest time-average gap in the three policies, and the gap shrinks with the increase of system requirement $\beta$. This shows the effectiveness of Algorithm 1 in searching feasible solutions with low payment to the original problem. Different from Fig. \ref{HCC_Payment}, the gap of the uniform policy is largest, which corresponds to the largest total payment, since it chooses the bids with the consideration in $\beta$ only and without considering $\alpha_l$. The results in Fig. \ref{SCC_Payment_2} is similar as in Fig. \ref{SCC_Payment_1}, except that the gap of our policy is slightly lower than that of the random policy when $\beta$ is large.

%\begin{figure}[!htb]
%\centering
%\includegraphics[scale = 0.22]{SCC_Payment_Gap_1}
%\caption{Time-Average Gap vs. System Requirement under Setting I.}
%\label{SCC_Payment_1}
%\end{figure}

%\begin{figure}[!htb]
%\centering
%\includegraphics[scale = 0.22]{SCC_Payment_Gap_2}
%\caption{Time-Average Gap vs. System Requirement under Setting II.}
%\label{SCC_Payment_2}
%\end{figure}
%\begin{figure}[t]
%\centering
%\begin{tabular}{ll}
%        {
%            \subfigure[Setting I]
%            {
%            \begin{minipage}[t]{0.2\textwidth}
%            \includegraphics[scale = 0.18]{SCC_Payment_Gap_1}
%            \end{minipage}
%            %\label{Rate5}
%            }
%        } &
%        {
%            \subfigure[Setting II]
%            {
%            \begin{minipage}[t]{0.2\textwidth}
%            \includegraphics[scale = 0.18]{SCC_Payment_Gap_2}
%            \end{minipage}
%            %\label{Rate10}
%            }
%        }
%    \end{tabular}
%\caption{Time-Average Gap vs. System Requirement under Two Settings.}
%\label{E2}
%\end{figure} %Performation Evaluation

\section{Conclusions}
In this paper, we studied how to minimize the total payment with the sensing robustness requirement in crowdsensing. We for the first time utilized chance constraints to model the sensing robustness requirement, and considered both the hard chance constraint case and soft chance constraint case. We reformulated the crowdsensing problem with a hard chance constraint into a solvable approximation problem. And we proposed a binary search algorithm based on Monte Carlo method, to obtain a feasible solution of the problem with soft chance constraints. We also theoretically analyzed the performance gap of our proposed policies in both two cases. In the future, we plan to consider more detail requirements in the sensing robustness rather than merely the minimum required participants.

% Acknowledgments here
%\ACKNOWLEDGMENT{%
% Enter the text of acknowledgments here
%}% Leave this (end of acknowledgment)

%\appendix
\begin{APPENDICES}
\section{Proof of Lemma 4}
We prove by contradiction. If the conclusion does not hold, there exists at least $t, \hat{t}\in\mathcal{T}$ ($t \neq \hat{t}$), which makes $f'_{tl}({\rho_t^\ast}^l) \neq f'_{\hat{t}l}({\rho_{\hat{t}}^\ast}^l)$. Without loss of generality, we assume that $f'_{tl}({\rho_t^\ast}^l) > f'_{\hat{t}l}({\rho_{\hat{t}}^\ast}^l)$ and $t < \hat{t}$. According to (\ref{eq31}) and (\ref{eq32}), we know that, for any $t\in\mathcal{T}$, $f'_{tl}(\cdot)$ is positive and strictly increasing in $\rho_t^l\in[0, 1]$. Thus, there exists a small value $\delta$ ($0 < \delta < \min \{{\rho_t^\ast}^l, 1 - {\rho_{\hat{t}}^\ast}^l\}$), such that
\begin{align}
\label{eq33}
f'_{tl}({\rho_t^\ast}^l - \delta) > f'_{\hat{t}l}({\rho_{\hat{t}}^\ast}^l + \delta)
\end{align}still holds.

Now, we can construct a new feasible solution of PB2.$l$, whose total payment is smaller than that with ${\rho^\ast}^l$ in fact. Consider the following sequence ${\rho_v^\ast}^l$:
\begin{displaymath}
{\rho_1^\ast}^l, ..., \rho_{t-1}^{\ast l}, {\rho_t^\ast}^l - \delta, \rho_{t+1}^{\ast l}, ..., \rho_{\hat{t}-1}^{\ast l}, {\rho_{\hat{t}}^\ast}^l + \delta, \rho_{\hat{t}+1}^{\ast l}, ..., \rho_T^{\ast l},
\end{displaymath}which simply substitutes ${\rho_t^\ast}^l$ and ${\rho_{\hat{t}}^\ast}^l$ in the optimal solution ${\rho^\ast}^l$ by ${\rho_t^\ast}^l - \delta$ and ${\rho_{\hat{t}}^\ast}^l + \delta$, respectively.

Since ${\rho^\ast}^l$ is feasible for PB2.$l$, then
\begin{align}
&{\rho_1^\ast}^l + \cdot\cdot\cdot + ({\rho_t^\ast}^l - \delta) + \cdot\cdot\cdot + ({\rho_{\hat{t}}^\ast}^l + \delta) + \cdot\cdot\cdot + \rho_T^{\ast l} \nonumber\\
&= {\rho_1^\ast}^l + \cdot\cdot\cdot + {\rho_t^\ast}^l + \cdot\cdot\cdot + {\rho_{\hat{t}}^\ast}^l + \cdot\cdot\cdot + \rho_T^{\ast l} \geq \gamma_l, \nonumber
\end{align}which directly implies that ${\rho_v^\ast}^l$ is also feasible for PB2.$l$.

Next, we show that the total payment with ${\rho_v^\ast}^l$ is smaller than that with ${\rho^\ast}^l$. Since the only differences between ${\rho_v^\ast}^l$ and ${\rho^\ast}^l$ are the $t$th component and $\hat{t}$th component, we have
\begin{align}
&F^l({\rho^\ast}^l) - F^l({\rho_v^\ast}^l) \nonumber \\
&= [f_{tl}({\rho_t^\ast}^l) + f_{\hat{t}l}({\rho_{\hat{t}}^\ast}^l)] - [f_{tl}({\rho_t^\ast}^l - \delta) + f_{\hat{t}l}({\rho_{\hat{t}}^\ast}^l + \delta)] \nonumber \\
&= [f_{tl}({\rho_t^\ast}^l) - f_{tl}({\rho_t^\ast}^l - \delta)] + [f_{\hat{t}l}({\rho_{\hat{t}}^\ast}^l) - f_{\hat{t}l}({\rho_{\hat{t}}^\ast}^l + \delta)] \nonumber \\
\label{eq34}
&\geq f'_{tl}({\rho_t^\ast}^l - \delta)\delta + f'_{\hat{t}l}({\rho_{\hat{t}}^\ast}^l + \delta)(-\delta) \\
\label{eq35}
&= \delta[f'_{tl}({\rho_t^\ast}^l - \delta) - f'_{\hat{t}l}({\rho_{\hat{t}}^\ast}^l + \delta)] > 0,
\end{align}where (\ref{eq34}) follows the convexity of $f_{tl}(\cdot)$ and $f_{\hat{t}l}(\cdot)$, and (\ref{eq35}) is due to (\ref{eq33}) and $\delta > 0$. Accordingly, the total payment with the optimal solution ${\rho^\ast}^l$, i.e., $F^l({\rho^\ast}^l)$, is larger than that with the feasible solution ${\rho_v^\ast}^l$, i.e., $F^l({\rho_v^\ast}^l)$, which obviously contradicts the optimality of ${\rho^\ast}^l$.

Therefore, for any $t, \hat{t}\in\mathcal{T}$ ($t \neq \hat{t}$), $f'_{tl}({\rho_t^\ast}^l) = f'_{\hat{t}l}({\rho_{\hat{t}}^\ast}^l)$ always holds. If $f'_{tl}({\rho_t^\ast}^l) \neq f'_{\hat{t}l}({\rho_{\hat{t}}^\ast}^l)$, we can repeat the above process in the similar way to construct a feasible solution with smaller total payment, which violates the optimality of the current solution. The proof thus ends.

\section{Proof of Lemma 5}%\ref{lemma6}}
This lemma can be proved by contradiction. First, we can draw a conclusion that there exists at least some $\tilde{t}\in\{1, ..., T\}$, such that ${\rho_{\tilde{t}}^\ast}^l(\hat{\gamma_l}) \geq {\rho_{\tilde{t}}^\ast}^l(\gamma_l)$ holds. Otherwise, for $\forall t = 1, ..., T$, ${\rho_t^\ast}^l(\hat{\gamma_l}) < {\rho_t^\ast}^l(\gamma_l)$ holds. Since ${\rho^\ast}^l(\hat{\gamma_l})$ is feasible for P2.$l$ with $\hat{\gamma_l}$, we obtain
\begin{displaymath}
\sum_{t=1}^T {\rho_t^\ast}^l(\hat{\gamma_l}) \geq \hat{\gamma_l} > \gamma_l,
\end{displaymath}which implies that ${\rho^\ast}^l(\hat{\gamma_l})$ is also feasible for PB2.$l$ with $\gamma_l$. And because $f_{tl}(\cdot)$ is strictly increasing in $[0, 1]$ for $t = 1, ..., T$, we have
\begin{displaymath}
F^l({\rho^\ast}^l(\hat{\gamma_l})) = \sum_{t=1}^L f_{tl}({\rho_t^\ast}^l(\hat{\gamma_l})) < \sum_{t=1}^L f_{tl}({\rho_t^\ast}^l(\gamma_l)) = F^l({\rho^\ast}^l(\gamma_l)).
\end{displaymath}According to the above, ${\rho^\ast}^l(\hat{\gamma_l})$ is feasible for PB2.$l$ with $\gamma_l$, whose total payment is smaller than the solution ${\rho^\ast}^l(\gamma_l)$. This result contradicts the optimality of ${\rho^\ast}^l(\gamma_l)$.

Hence, ${\rho_{\tilde{t}}^\ast}^l(\hat{\gamma_l}) \geq {\rho_{\tilde{t}}^\ast}^l(\gamma_l)$ holds for some $\tilde{t}\in\{1, ..., T\}$. This will result in the fact that ${\rho_{t}^\ast}^l(\hat{\gamma_l}) \geq {\rho_{t}^\ast}^l(\gamma_l)$ holds for any $t\in\mathcal{T}$. Otherwise, there exists some $\acute{t}\in\mathcal{T}$, such that ${\rho_{\acute{t}}^\ast}^l(\hat{\gamma_l}) < {\rho_{\acute{t}}^\ast}^l(\gamma_l)$ holds. Since $f'_{tl}(\cdot)$ is strictly increasing in $[0, 1]$ for $t = 1, ..., T$, we obtain
\begin{align}
\label{eq36}
f'_{\tilde{t}l}({\rho_{\tilde{t}}^\ast}^l(\hat{\gamma_l})) > f'_{\tilde{t}l}({\rho_{\tilde{t}}^\ast}^l(\gamma_l)).
\end{align}According to Lemma \ref{lemma5}, we have
\begin{align}
f'_{\tilde{t}l}({\rho_{\tilde{t}}^\ast}^l(\hat{\gamma_l})) = f'_{\acute{t}l}({\rho_{\acute{t}}^\ast}^l(\hat{\gamma_l})), \nonumber\\
\label{eq37}
f'_{\tilde{t}l}({\rho_{\tilde{t}}^\ast}^l(\gamma_l)) = f'_{\acute{t}l}({\rho_{\acute{t}}^\ast}^l(\gamma_l)).
\end{align}Based on (\ref{eq36}) and (\ref{eq37}), we have
\begin{align}
\label{eq38}
f'_{\acute{t}l}({\rho_{\acute{t}}^\ast}^l(\hat{\gamma_l})) > f'_{\acute{t}l}({\rho_{\acute{t}}^\ast}^l(\gamma_l)),
\end{align}which implies ${\rho_{\acute{t}}^\ast}^l(\hat{\gamma_l}) > {\rho_{\acute{t}}^\ast}^l(\gamma_l)$, due to the strictly increasing property of $f'_{\acute{t}l}(\cdot)$. Thus, the assumption does not holds. In other words, for any $t\in\mathcal{T}$, ${\rho_{t}^\ast}^l(\hat{\gamma_l}) \geq {\rho_{t}^\ast}^l(\gamma_l)$.

Next, we prove that the equality cannot be maintained for ${\rho_{t}^\ast}^l(\hat{\gamma_l}) \geq {\rho_{t}^\ast}^l(\gamma_l)$, for any $t\in\mathcal{T}$. Similar to the proof of ${\rho_{\acute{t}}^\ast}^l(\hat{\gamma_l}) \geq {\rho_{\acute{t}}^\ast}^l(\gamma_l)$ in the above, if ${\rho_{\tilde{t}}^\ast}^l(\hat{\gamma_l}) = {\rho_{\tilde{t}}^\ast}^l(\gamma_l)$ holds for some $\tilde{t}\in\mathcal{T}$, then it holds for any $t\in\mathcal{T}$. This utilizes the conclusion of Lemma \ref{lemma5}, and the strictly increasing property of $f'_{tl}(\cdot)$ for any $t\in\mathcal{T}$.

Further, it is not difficult to verify that PB2.$l$ is a convex problem with strong duality. According to the KKT conditions, we have
\begin{align}
\label{eq39}
\sum_{t=1}^T f'_{tl}({\rho_t^\ast}^l(\gamma_l)) - \lambda^\ast(\gamma_l) T = 0, \\
\label{eq40}
\lambda^\ast(\gamma_l) [\sum_{t=1}^T {\rho_t^\ast}^l(\gamma_l) - \gamma_l] = 0,
\end{align}where $\lambda^\ast(\gamma_l)$ is the optimal solution of the dual problem corresponding to PB2.$l$ with $\gamma_l$. Since $f'_{tl}({\rho_t^\ast}^l(\gamma_l))$ is positive for $t = 1, ..., T$, $\lambda^\ast(\gamma_l)$ is also positive according to Eq. (\ref{eq39}). Then, from Eq. (\ref{eq40}), we know that
\begin{align}
\label{eq41}
\sum_{t=1}^T {\rho_t^\ast}^l(\gamma_l) - \gamma_l = 0,
\end{align}which also holds for the substitution of $\hat{\gamma_l}$ for $\gamma_l$. Therefore, if ${\rho_t^\ast}^l(\gamma_l) = {\rho_t^\ast}^l(\hat{\gamma_l})$ for $t = 1, ..., T$, due to Eq. (\ref{eq41}), we obtain $\gamma_l = \hat{\gamma_l}$, which violates the initial assumption.

To summarize, if $\hat{\gamma_l} > \gamma_l$, then ${\rho_t^\ast}^l(\hat{\gamma_l}) > {\rho_t^\ast}^l(\gamma_l)$ holds for any $t\in\mathcal{T}$. The lemma is thus proved.

\section{Proof of Theorem 4}%\ref{theorem3}}
We begin with some explanations of $h^l({\rho}^l)$. Let $k_l$ be the integer that is the nearest interger to $T\alpha_l$ and no less than $T\alpha_l$, i.e.,
\begin{align}
&k_l := \left\{
     \begin{array}{l l}
           T\alpha_l & \mbox{if $T\alpha_l$ is an integer}, \\
           {[T\alpha_l]}+1 & \mbox{otherwise}, \nonumber
     \end{array}\right.
\end{align}where $[\cdot]$ is the round operation. Since $\alpha_l\in(0, 1]$, we have $k_l\in\{1, ..., T\}$. Then, $h^l({\rho}^l)$ is equivalent to
\begin{displaymath}
h^l({\rho}^l) = \mbox{Pr}\left\{\sum_{t=1}^T \bar{I}_t^l(x_t^l) \geq k_l\right\}.
\end{displaymath}Recall that $A_t^l$ denotes the event of $x_t^l = r_t^l$ for $t = 1, ..., T$ and $l = 1, ..., L$, whose success probability is $\rho_t^l$. According to the definition of $\bar{I}_t^l(x_t^l)$ in Eq. (\ref{eq24}), $\sum_{t=1}^T \bar{I}_t^l(x_t^l)$ is equal to the number of success events from the $T$ independent events, i.e., $A_1^l, ..., A_T^l$. Therefore, $h^l({\rho}^l)$ is equal to the probability that at least $k_l$ events happen in the $T$ events, where the success probability of the $t$th event is $\rho_t^l$, i.e., $\mbox{Pr}\{A_t^l\} = \rho_t^l$. In the following, we focus on $h^l({\rho^\ast}^l(\gamma_l))$, where ${\rho^\ast}^l(\gamma_l) := \{{\rho_t^\ast}^l(\gamma_l)\}_{t = 1, ..., T}$ is an optimal solution of PB2.$l$ with $\gamma_l$.

Based on Lemma \ref{lemma6}, if $\gamma_l$ increases to $\tilde{\gamma_l}$, the corresponding optimal solution also increases, i.e., for any $t = 1, ..., T$, ${\rho_t^\ast}^l(\tilde{\gamma_l})$ is larger than ${\rho_t^\ast}^l(\gamma_l)$. This means that the success probability of any event $A_t^l$ ($t = 1, ..., T$) is increased. In the following, by mathematical induction on the number of events (e.g., $i$) with increased success probabilities, we prove that $h^l({\rho^\ast}^l(\tilde{\gamma_l}))$ also becomes larger than $h^l({\rho^\ast}^l(\gamma_l))$ when the success probabilities of all events are increased.

For the base case, i.e., $i = 1$, we first show that the conclusion holds, when there is only one event whose success probability increased. Since the $T$ events are independent with each other, without loss of generality, assume that only ${\rho_1^\ast}^l(\gamma_l)$ is increased to ${\rho_1^\ast}^l(\tilde{\gamma_l})$, while ${\rho_2^\ast}^l(\gamma_l), ..., {\rho_T^\ast}^l(\gamma_l)$ keep unchanged. Let an experiment be a run of the $T$ events. Consider the two experiments with the two following success probability sequences, respectively: \\
1) ${\rho_1^\ast}^l(\gamma_l), {\rho_2^\ast}^l(\gamma_l), ..., {\rho_T^\ast}^l(\gamma_l)$; \\
2) ${\rho_1^\ast}^l(\tilde{\gamma_l}), {\rho_2^\ast}^l(\gamma_l), ..., {\rho_T^\ast}^l(\gamma_l)$.\\
We now proceed to compare the outputs of the two above experiments, where the output is defined as the number of total success events, i.e., $\sum_{t=1}^T \bar{I}_t^l(x_t^l)$. We say an experiment succeeds only if $\sum_{t=1}^T \bar{I}_t^l(x_t^l)$ is no less than $k_l$.

Let $k^-$ be the number of success events from $\{A_t^l\}_{t = 2, ..., T}$. Based on the value of $k^-$, we divide the output into three cases as follows. If $k^- \geq k_l$, which corresponds to the case when at least $k_l$ events from $\{A_t^l\}_{t = 2, ..., T}$ happen, then $\sum_{t=1}^T \bar{I}_t^l(x_t^l) \geq k_l$ holds, no matter whether the event $A_1^l$ happens or not. If $k^- < k_l - 1$, corresponding to the case when the number of success events from $\{A_t^l\}_{t = 2, ..., T}$ is less than $k_l - 1$, $\sum_{t=1}^T \bar{I}_t^l(x_t^l)$ will be less than $k_l$, no matter whether the event $A_1^l$ happens or not. In the above two cases, the outputs of the two experiments are identical. If $k^- = k_l -1$, which corresponds to the case when there are exactly $k_l - 1$ success events from $\{A_t^l\}_{t = 2, ..., T}$, the output will be determined by the result of the event $A_1^l$. Since ${\rho_1^\ast}^l(\tilde{\gamma_l})$ is larger than ${\rho_1^\ast}^l(\gamma_l)$, the probability of $\sum_{t=1}^T \bar{I}_t^l(x_t^l) = k_l$ in the second experiment is higher than that in the first experiment. To sum, the success probability of the second experiment is higher than that of the first one, which directly implies $h^l({\rho^\ast}^l(\tilde{\gamma_l})) > h^l({\rho^\ast}^l(\gamma_l))$.

Next, we assume that the conclusion holds for $i = n$ ($n \leq T - 1$). Without loss of generality, suppose that ${\rho_t^\ast}^l(\gamma_l)$ is increased to ${\rho_t^\ast}^l(\tilde{\gamma_l})$, $t = 1, ..., n$, while ${\rho_t^\ast}^l(\gamma_l)$ is unchanged for $t = n+1, ..., T$. Let ${\rho^\ast_v}^l$ be the corresponding changed success probability sequence, i.e.,
\begin{displaymath}{\rho^\ast_v}^l := \{{\rho_1^\ast}^l(\tilde{\gamma_l}), ..., {\rho_n^\ast}^l(\tilde{\gamma_l}), \rho_{n+1}^{\ast l}(\gamma_l), \rho_{n+2}^{\ast l}(\gamma_l), ..., \rho_T^{\ast l}(\gamma_l)\}.
\end{displaymath}Under our assumption, we obtain
\begin{align}
\label{eq42}
h^l({\rho^\ast_v}^l) > h^l({\rho^\ast}^l(\gamma_l)).
\end{align}We then consider the case when we increase one more success probability, e.g., $\rho_{n+1}^{\ast l}(\gamma_l)$ is increased to $\rho_{n+1}^{\ast l}(\tilde{\gamma_l})$. And let $\rho^{\ast l}_{v^+}$ denote the changed probability sequence based on ${\rho^\ast_v}^l$, i.e.,
\begin{displaymath}\rho^{\ast l}_{v^+} := \{{\rho_1^\ast}^l(\tilde{\gamma_l}), ..., {\rho_n^\ast}^l(\tilde{\gamma_l}), {\rho_{n+1}^{\ast l}}(\tilde{\gamma_l}), \rho_{n+2}^{\ast l}(\gamma_l), ..., {\rho_T^\ast}^l(\gamma_l)\}.
\end{displaymath}

Compared to ${\rho^\ast_v}^l$, the only difference is that $\rho_{n+1}^{\ast l}(\gamma_l)$ is substituted by $\rho_{n+1}^{\ast l}(\tilde{\gamma_l})$ in $\rho^{\ast l}_{v^+}$. Since $\rho_{n+1}^{\ast l}(\tilde{\gamma_l})$ is larger than $\rho_{n+1}^{\ast l}(\gamma_l)$, we can apply the similar process as in the base case to prove that $h^l(\rho^{\ast l}_{v^+}) > h^l({\rho^\ast_v}^l)$ holds. Due to (\ref{eq42}), we have $h^l(\rho^{\ast l}_{v^+}) > h^l(\rho^{\ast l}(\gamma_l))$, which implies that the conclusion holds for $i = n + 1$.

Therefore, the conclusion comes into existence for any $i = 1, ..., T$. The proof is thus concluded.

\section{Proof of Lemma 6}%\ref{lemma7}}
The proof consists of two steps. First, we show that For any feasible solution $\rho = \{\rho_t^l\}_{t\in\mathcal{T}, l\in\mathcal{L}}$ to PB2, we can construct a feasible solution $\hat{\rho} = \{\hat{\rho}^l\}_{l\in\mathcal{L}}$ to PB4, where
\begin{displaymath}\hat{\rho}^l = \frac{\sum_{t=1}^T \rho_t^l}{T}, \forall l\in\mathcal{L}.\end{displaymath}
And $\hat{F}(\hat{\rho}) \leq F(\rho)$ holds. For the feasibility, we only need to prove that the constructed $\hat{\rho}$ can satisfy (\ref{eq46}). Since $\rho$ is a feasible solution of PB2, i.e.,
\begin{align}
\label{eq47}
\sum_{t=1}^T \rho_t^l \geq \gamma_l,
\end{align}which utilizes the fact that $\mathbb{E}\{\bar{I}_t^l(x_t^l)\} = \rho_t^l$. According to (\ref{eq47}) and the definition of $\hat{\rho}$, $\hat{\rho}$ is thus feasible for PB4.

Next, recall that we have proved that function $xb_l(x)$ is convex in $x\in[0, 1]$ in Lemma \ref{lemma3}. Therefore, we have
\begin{align}
\label{eq48}
\hat{\rho}^l b_l(\hat{\rho}^l) \leq \sum_{t=1}^T \frac{1}{T} \rho_t^l b_l(\rho_t^l).
\end{align}Multiplying by $Tr^l$ in both sides and summing over $l = 1, ..., L$ in ($\ref{eq48}$), we have
\begin{displaymath}
\sum_{l=1}^L T r^l \hat{\rho}^l b_l(\hat{\rho}^l) \leq \sum_{l=1}^L r^l \sum_{t=1}^T \rho_t^l b_l(\rho_t^l),
\end{displaymath}which results in $\hat{F}(\hat{\rho}) \leq F(\rho)$. The first step is completed.

Second, let $\hat{\rho}^{\ast} = \{\hat{\rho}^l_{\ast}\}_{l\in\mathcal{L}}$ be the optimal solution of PB4. We prove that we can obtain a feasible solution $\rho^{\ast} = \{{\rho_t^l}^{\ast}\}_{t\in\mathcal{T}, l\in\mathcal{L}}$ to PB2, where ${\rho_t^l}^{\ast} = \hat{\rho}^l_\ast, \forall t\in\mathcal{T},$ for each $l\in\mathcal{L}$. Moreover, $\hat{F}(\hat{\rho}^\ast) = F(\rho^\ast)$ holds. The feasibility can be verified simply by the definition of $\rho^{\ast}$ and the feasibility of $\hat{\rho}^l_\ast$ to (\ref{eq46}). As to the equivalence, by definition, the objective value of PB2 for solution $\rho^{\ast}$ is
\begin{align}
&F(\rho^{\ast}) = \sum_{t=1}^T \sum_{l=1}^L {\rho_t^l}^{\ast} r^l b_l({\rho_t^l}^{\ast}) \nonumber \\
&\quad\quad\quad = \sum_{t=1}^T \sum_{l=1}^L \hat{\rho}^l_\ast r^l b_l(\hat{\rho}^l_\ast)
= T \sum_{l=1}^L \hat{\rho}^l_\ast r^l b_l(\hat{\rho}^l_\ast) = F(\hat{\rho}^{\ast}). \nonumber
\end{align}Based on the above steps, the lemma is thus proved.
\end{APPENDICES}

% References here (outcomment the appropriate case)

% CASE 1: BiBTeX used to constantly update the references
%   (while the paper is being written).
%\bibliographystyle{ijocv081} % outcomment this and next line in Case 1
%\bibliography{<your bib file(s)>} % if more than one, comma separated

% CASE 2: BiBTeX used to generate mypaper.bbl (to be further fine tuned)
%\input{mypaper.bbl} % outcomment this line in Case 2
%References
%\bibliographystyle{IEEEtran}
\bibliographystyle{ormsv080}

\end{document}